\documentclass[utf8]{aa} 
\pdfoutput=1 

\usepackage{xcolor}
\usepackage{txfonts} 
\usepackage{physics} 
\usepackage{graphicx} 
\usepackage[alsoload=astro]{siunitx} 

\usepackage{hyperref} 
\hypersetup{
    colorlinks   = true, 
    urlcolor     = blue, 
    linkcolor    = blue, 
    citecolor   = blue 
}

\defcitealias{Henriques2015}{H15}
\defcitealias{Linke2020}{L20}
\defcitealias{Lagos2012}{L12}

\makeatletter
\renewcommand*\aa@pageof{, page \thepage{} of \pageref*{LastPage}}
\makeatother

\begin{document} 
       
    \title{KiDS+VIKING+GAMA: Testing semi-analytic models of galaxy evolution with galaxy-galaxy-galaxy lensing}
    
    \titlerunning{KiDS+VIKING+GAMA: Testing SAMs with galaxy-galaxy-galaxy-lensing}
    
    \author{Laila Linke \inst{1}
        \and
        Patrick Simon \inst{1}
        \and
        Peter Schneider \inst{1}
        \and
        Thomas Erben \inst{1}
         \and
         Daniel J. Farrow \inst{2}
         \and 
        Catherine Heymans \inst{3,4}
        \and
        \\Hendrik Hildebrandt \inst{4}
        \and
        Andrew M. Hopkins \inst{5}
        \and
        Arun Kannawadi\inst{6, 7}
        \and
        Nicola R. Napolitano \inst{8}
        \and
        Crist\'{o}bal Sif\'{o}n \inst{9}
        \and
        \\Angus H. Wright \inst{4}
    }
    
    \authorrunning{L.~Linke et al.}
    
    \institute{Argelander-Institut f\"ur Astronomie, Rheinische Friedrich-Wilhelms-Universit\"at, Auf dem H\"ugel 71, 53121 Bonn, Germany\\
        \email{llinke@astro.uni-bonn.de}
        \and 
        Max-Planck-Institut für extraterrestrische Physik, Giessenbachstrasse 1, 85748 Garching, Germany
        \and
         Institute for Astronomy, University of Edinburgh, Royal Observatory, Blackford Hill, Edinburgh, EH9 3HJ, UK
        \and Ruhr-Universit\"at Bochum, Astronomisches Institut, German Centre for Cosmological Lensing (GCCL), Universit\"atsstr. 150, 44801 Bochum, Germany
          \and
        Australian Astronomical Optics, Macquarie University, 105 Delhi Rd, North Ryde NSW 2113, Australia
        \and
        Department of Astrophysical Sciences, Princeton University, 4 Ivy Lane, Princeton, NJ 08544, USA
         \and
         Leiden Observatory, Leiden University, P.O.Box 9513, 2300RA Leiden, The Netherlands 
        \and
        School for Physics and Astronomy, Sun Yat-sen University, Guangzhou 519082, Zhuhai Campus, China
        \and
        Instituto de F\'{\i}sica, Pontificia Universidad Cat\'{o}lica de Valpara\'{\i}so, Casilla 4059, Valpara\'{\i}so, Chile    }
    
    \date{Received 6 May 2020; accepted 9 June 2020}
    
    \abstract{Several semi-analytic models (SAMs) try to explain how galaxies form, evolve, and interact inside the dark matter large-scale structure. These SAMs can be tested by comparing their predictions for galaxy-galaxy-galaxy lensing (G3L), which is weak gravitational lensing around galaxy pairs, with observations. }{We evaluate the SAMs by \citet[{H15}]{Henriques2015} and by \citet[L12]{Lagos2012}, which were implemented in the Millennium Run, by comparing their predictions for G3L to observations at smaller scales than previous studies and also for pairs of lens galaxies from different populations.}{We compared the G3L signal predicted by the SAMs to measurements in the overlap of the Galaxy And Mass Assembly survey (GAMA), the Kilo-Degree Survey (KiDS), and the VISTA Kilo-degree Infrared Galaxy survey (VIKING) by splitting lens galaxies into two colour and five stellar-mass samples. Using an improved G3L estimator, we measured the three-point correlation of the matter distribution with `mixed lens pairs' with galaxies from different samples, and with `unmixed lens pairs' with galaxies from the same sample.}{Predictions by the \citetalias{Henriques2015} SAM for the G3L signal agree with the observations for all colour-selected samples and all but one stellar-mass-selected sample with 95\% confidence. Deviations occur for lenses with stellar masses below $9.5\,h^{-2}\,\mathrm{M}_{\odot}$ at scales below $0.2\,h^{-1}\,\mathrm{Mpc}$. Predictions by the \citetalias{Lagos2012} SAM for stellar-mass selected samples and red galaxies are significantly higher than observed, while the predicted signal for blue galaxy pairs is too low.}{The \citetalias{Lagos2012} SAM predicts more pairs of low stellar mass and red galaxies than the \citetalias{Henriques2015} SAM and the observations, as well as fewer pairs of blue galaxies. This difference increases towards the centre of the galaxies' host halos. Likely explanations are different treatments of environmental effects by the SAMs and different models of the initial mass function. We conclude that G3L provides a stringent test for models of galaxy formation and evolution.}
    
    \keywords{Gravitational lensing: weak -- cosmology: observations -- large-scale structure -- Galaxies: evolution}
    
    \maketitle
    %
    \section{Introduction}
    \label{sec:intro}
    One important goal of extragalactic astronomy and cosmology is understanding galaxy formation and evolution. The following two different types of simulations try to reproduce the observed galaxy and matter distribution: full hydrodynamical simulations \citep[e.g.][]{Illustris2014, Eagle2015, Kaviraj2017, Nelson2019} and dark-matter-only $N$-body simulations with galaxies inserted according to semi-analytic models (SAMs) of galaxy formation and evolution.
    
    Multiple SAMs, with different assumptions on small-scale physics, such as the gas cooling time, the star formation rate, or supernovae feedback, have been proposed \citep[e.g.][]{Bower2006, Guo2011, Lagos2012, Henriques2015}. These models must be assessed by comparing their predictions to observations of galaxy statistics. Previous tests of SAMs included galaxy-galaxy lensing \citep[GGL; e.g.][]{Saghiha2017} and galaxy clustering \citep[e.g.][]{Henriques2017}.
    
    A more sensitive test than GGL is comparing the galaxy-galaxy-galaxy lensing (G3L) signal predicted by the SAMs to observations. The G3L effect, which was first discussed by \citet{Schneider2005}, describes the weak gravitational lensing of pairs of background galaxies around foreground galaxies (lens-shear-shear correlation) and of individual background galaxies around pairs of foreground galaxies (lens-lens-shear correlation). Unlike GGL or galaxy clustering, G3L depends on the galaxy-matter three-point correlation and the halo occupation distribution of galaxy pairs. In principle, it also depends on the ellipticity of dark matter halos as well as misalignments between the galaxy and matter distribution because the galaxy pair orientation introduces a preferred direction.
    
    The lens-lens-shear correlation was measured for lens pairs separated by several megaparsecs (Mpc) in order to detect inter-cluster filaments \citep{Mead2010, Clampitt2016, Epps2017, Xia2020}. However, for the evaluation of SAMs, it is more suitable to study the correlation at smaller, sub-Mpc scales. At these scales, the G3L signal is more sensitive to the small-scale physics that vary between different SAMs, because it depends primarily on galaxy pairs with galaxies in the same dark matter halo. For lens pairs with galaxies of a similar stellar mass or colour, the small-scale lens-lens-shear correlation was determined by \citet{Simon2008} in the Red Cluster Sequence survey \citep{Hildebrandt2016} and \citet{Simon2013} in the Canada-France-Hawaii Telescope Lensing Survey (CFHTLenS; \citealp{Heymans2012}). The G3L measured in CFHTLenS was compared to predictions by multiple SAMs that were implemented in the Millennium Run \citep[MR;][]{Springel2005} by \citet{Saghiha2017} and \citet{Simon2019}. They demonstrate that G3L is more effective in evaluating SAMs than GGL and that the SAM by \citet[H15 hereafter]{Henriques2015} is in agreement with the observations in CFHTLenS, while the SAM by \citet[L12 hereafter]{Lagos2012} predicts G3L signals that are too large.

Nonetheless, these previous measurements of G3L at small scales only used photometric data with imprecise redshift estimates for the lens galaxies. Therefore, lens galaxy pairs with galaxies separated along the line-of-sight (chance pairs) were treated the same as lens galaxy pairs with galaxies close to each other (true pairs). As the G3L signal of chance pairs is much weaker, this lowers the signal-to-noise ratio (S/N). 
    
    However, \citet[][L20 hereafter]{Linke2020} demonstrate that the S/N could be improved substantially by weighting each lens galaxy pair according to the line-of-sight separation between its galaxies to reduce the impact of chance pairs. We used this improved estimator to test the \citetalias{Henriques2015} and the \citetalias{Lagos2012} SAMs with state-of-the-art observational data, consisting of the photometric Kilo-Degree Survey (KiDS) and VISTA Kilo-degree Infrared Galaxy survey (VIKING) as well as the spectroscopic Galaxy And Mass Assembly survey (GAMA). We used the shapes of galaxies observed by KiDS as shear estimates, while GAMA provides lens galaxies with precise spectroscopic redshifts. These spectroscopic redshifts allowed us to employ the redshift weighting suggested by \citetalias{Linke2020}. Furthermore, we extended the angular range at which we measured the G3L signal to lower scales with the adaptive binning scheme for the G3L three-point correlation function proposed by \citetalias{Linke2020}. Thereby, we could assess the SAMs deeper inside dark-matter halos.
    
    As of now, the lens-lens-shear correlation has only been measured for lens pairs with galaxies from the same colour or stellar-mass sample (unmixed lens pairs) and not for lens pairs with galaxies from different samples (mixed lens pairs). However, comparing the measurements for G3L with mixed pairs is a compelling new test of SAMs, because this signal depends on the correlation of different galaxy populations inside halos. For example, the mixed pair G3L signal would be higher for two fully correlated galaxy populations than for two uncorrelated populations, while the GGL signal would stay the same. Therefore, we can assess the predictions of SAMs for the correlation between different galaxy populations with the G3L of mixed lens pairs. Accordingly, we measure not only the G3L signal for lens pairs from the same population but also the signal for mixed lens pairs, with galaxies from different colour- or stellar-mass samples. 
    
    This paper is structured as follows: In Sect.~\ref{sec:theory} we review the basics of G3L and introduce the third-order aperture statistics, which are the G3L observables throughout this work. Section~\ref{sec:methods} discusses our estimators for the three-point correlation function and the aperture statistics. The SAMs and the creation of the simulated and observational data sets are described in Sect.~\ref{sec:data}. We present the G3L signals measured in the observation and the simulation in Sect.~\ref{sec:results} and discuss our findings in Sect.~\ref{sec:discussion}.
    
    Throughout this paper we assume a flat $\Lambda$CDM cosmology with matter density $\Omega_\textrm{m}=0.25$, baryon density $\Omega_\textrm{b}= 0.045$, dark energy density $\Omega_\Lambda= 0.75$, Hubble constant $H_0= 73 \,\textrm{km}\,\textrm{s}^{-1}\,\textrm{Mpc}^{-1}$ and power spectrum normalisation $\sigma_8=0.9$. These parameters were used in the creation of the MR and differ from more recent constraints \citep[e.g.][]{Planck2018}. However, weak gravitational lensing is most sensitive to the combination of the matter density and the power spectrum normalisation $S_8=\sigma_8\,\sqrt{\Omega_\textrm{m}/0.3}$. This parameter is almost the same in the MR and the most recent Planck measurements; it is $S_{8, \mathrm{MR}}=0.822$ for the MR and $S_{8, \mathrm{Planck}}=0.825\pm 0.011$ in \citet{Planck2018}.
    
    
    \section{Theory of galaxy-galaxy-galaxy-lensing}
    \label{sec:theory}
    \begin{figure}
        \centering
        \resizebox{\hsize}{!}{\includegraphics{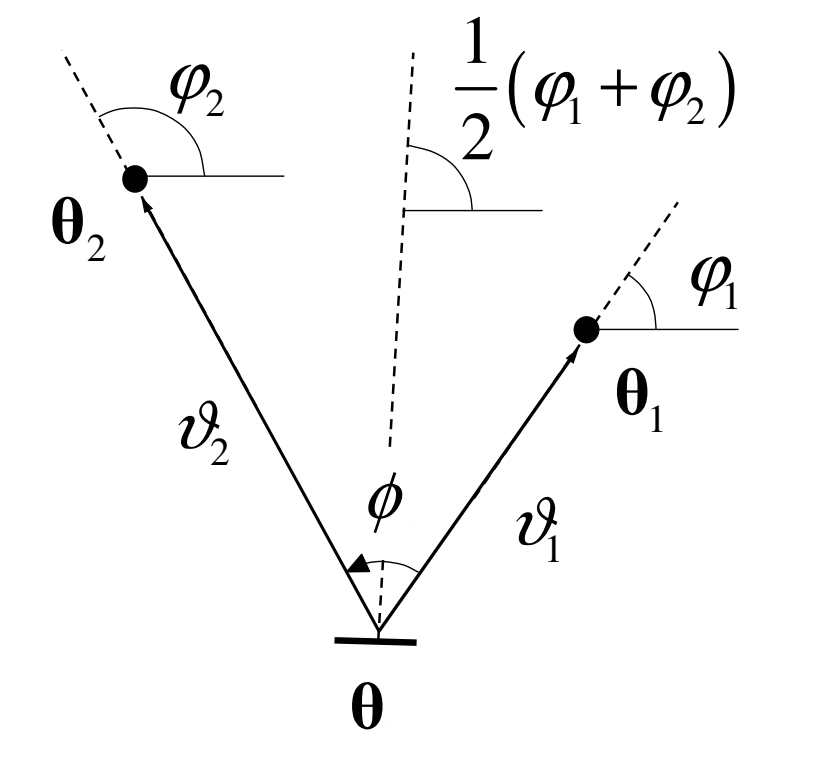}}
        \caption{Geometry of a G3L configuration with one source and two lens galaxies; adapted from \citet{Schneider2005}. The two lens galaxies are at angular positions $\va*{\theta}_1$ and $\va*{\theta}_2$ on the sky; the source galaxy is at $\va*{\theta}$. The separation vectors $\va*{\vartheta}_1$ and $\va*{\vartheta}_2$ of the lenses from the source  have lengths $\vartheta_1$ and $\vartheta_2$, as well as polar angles $\varphi_1$ and $\varphi_2$. The angle between $\va*{\vartheta}_1$ and $\va*{\vartheta}_2$ is the opening angle $\phi$. The tangential shear of the source galaxy is measured with respect to the dashed line, which is the bisector of $\phi$.}
        \label{fig:G3L}
    \end{figure}
    
    G3L is a weak gravitational lensing effect \citep[see e.g.][]{Bartelmann2001}. It comprises the correlation of individual lens galaxies with the shear of source galaxy pairs, as well as the correlation of lens galaxy pairs with the shear of individual source galaxies. Here, we study the second effect, the lens-lens-shear correlation. Figure~\ref{fig:G3L} illustrates the geometric configuration of lens and source galaxies for this correlation.
    
    \subsection{Three-point correlation function}
    \label{sec:theory:three-point function}
    The main observable for the lens-lens-shear correlation is the three-point correlation function $\tilde{\mathcal{G}}$. This function correlates the projected lens galaxy number density and the tangential gravitational lensing shear $\gamma_\textrm{t}$, measured with respect to the bisector of the angle $\phi$ between the lens-source separations $\va*{\vartheta}_1$ and $\va*{\vartheta}_2$ (see Fig.~\ref{fig:G3L}). For unmixed lens pairs, whose galaxies have the projected number density $N(\va*{\vartheta})$, $\tilde{\mathcal{G}}$ is
    \begin{equation}
    \label{eq:Gtilde}
    \tilde{\mathcal{G}}({\va*{\vartheta}_1}, {\va*{\vartheta}_2})=\dfrac{1}{\overline{N}^2}\,\expval{ N(\va*{\theta}+\va*{\vartheta}_1)\, N(\va*{\theta}+\va*{\vartheta}_2) \, \gamma_\textrm{t}(\va*{\theta})}\;,
    \end{equation}
    where $\overline{N}$ is the mean number density of lenses. For mixed lens pairs, where the lenses are from samples with number densities $N_1$ and $N_2$, the correlation function is
    \begin{equation}
    \label{eq:Gtilde_2samples}
    \tilde{\mathcal{G}}({\va*{\vartheta}_1}, {\va*{\vartheta}_2})=\frac{1}{\overline{N}_1\overline{N}_2}\,\expval{ N_1(\va*{\theta}+\va*{\vartheta}_1)\, N_2(\va*{\theta}+\va*{\vartheta}_2) \, \gamma_\textrm{t}(\va*{\theta})}\;.
    \end{equation}
    
    Instead of measuring $\tilde{\mathcal{G}}$, we estimated a redshift-weighted correlation function $\tilde{\mathcal{G}}_Z$, which includes a redshift weighting function $Z$, which depends on the redshift difference $\Delta z_{12}=z_1-z_2$ of the lenses in a pair. We chose the redshift weighting such that it is large for small $\Delta z_{12}$ and vanishes for large $\Delta z_{12}$. It, therefore, weights true pairs with small redshift differences higher than chance pairs with large redshift differences and increases the S/N \citepalias{Linke2020}. To define the redshift-weighted correlation function $\tilde{\mathcal{G}}_Z$, we used that the projected lens number densities $N_{1,2}$ are related to the three-dimensional number densities $n_{1,2}(\va*{\theta}, z)$ at redshift $z$ by the selection functions $\nu_{1,2}(z)$,
    \begin{equation}
        \label{eq:projectednumberdensity}
        N_{1,2}(\va*{\theta}) = \int \dd{z}\; \nu_{1,2}(z)\, n_{1,2}(\va*{\theta}, z)\;.
    \end{equation}
    The selection functions give the fraction of galaxies at redshift $z$ included in the lens sample. For a flux-limited galaxy sample, this corresponds to the fraction of galaxies brighter than the magnitude limit. The selection functions $\nu_{1,2}(z)$ are related to the galaxies' redshift distributions $p_{1,2}(z)$ by
    \begin{equation}
        \nu_{1,2}(z) = p_{1,2}(z)\, \frac{\int_A \dd[2]{\theta}\; N_{1,2}(\va*{\theta})}{\int_A \dd[2]{\theta}\; n_{1,2}(\va*{\theta}, z)}\;.
    \end{equation}
    
    With Eq.~\eqref{eq:projectednumberdensity}, the redshift-weighted correlation function is
    \begin{align}
    \label{eq:Gtilde redshiftweighted}
    &\notag \tilde{\mathcal{G}}_Z(\va*{\vartheta}_1, \va*{\vartheta}_2) \\
    &\notag= \left[\int_0^\infty \dd{z_1} \int_0^\infty \dd{z_2} \, \nu_1(z_1)\, \nu_2(z_2) Z(\Delta z_{12})\, \bar{n}_1(z_1)\, \bar{n}_2(z_2)\right]^{-1}\\
    &\quad \times \int_0^\infty \dd{z_1} \int_0^\infty \dd{z_2} \, \nu_1(z_1)\,\nu_2(z_2)\, Z(\Delta z_{12}) \\
    &\notag \quad \times \expval{n_1(\va*{\theta} + \va*{\vartheta}_1, z_1)\,n_2(\va*{\theta} + \va*{\vartheta}_2, z_2) \, \gamma_\textrm{t}(\va*{\theta})}\;,
    \end{align}
    where $\Delta z_{12} := z_1 - z_2$ is the redshift difference between lenses in a pair and 
    \begin{equation}
        \bar{n}_{1,2}(z) = \expval{n_{1,2}(\va*{\theta}, z)} = \frac{1}{A} \int_A \dd[2]{\theta}\; n_{1,2}(\va*{\theta}, z)\;. 
    \end{equation}
    
    Additionally, we also measured the physical correlation function $\tilde{\mathcal{G}}_\textrm{phys}(\va*{r}_1, \va*{r}_2)$, which gives the projected excess mass around lens pairs with physical lens-source separations $\va*{r}_{1}$ and $\va*{r}_2$ projected on a plane midway between the lenses. To find $\tilde{\mathcal{G}}_\textrm{phys}$, instead of averaging the tangential shear $\gamma_\textrm{t}$, we averaged the projected excess mass density $\Delta \Sigma$ around lens pairs with
    \begin{align}
    \label{eq:Gtilde Phys}
    &\notag\tilde{\mathcal{G}}_\textrm{phys}(\va*{r}_1, \va*{r}_2) \\ 
    &\notag=\left[\int_{0}^{\infty} \dd{z_1} \, \int_{0}^{\infty} \dd{z_2}\; \nu_1(z_1)\, \nu_2(z_2)\,  Z(\Delta z_{12})\, \bar{n}_1(z_1)\, \bar{n}_2(z_2)\right]^{-1}\\
    & \quad \times \int \dd{z_1}\, \int \dd{z_2}\; \nu_1(z_1)\, \nu_2(z_2) \, Z(\Delta z_{12})\\
    &\notag \quad\times \expval{n_1\left(\va*{\theta}+D^{-1}_{12} \,\va{r}_1, z_1\right)\, n_2\left(\va*{\theta}+D^{-1}_{12}\,\va{r}_2 , z_2 \right)\, \Delta \Sigma(\va*{\theta}, z_{12})}\;,
    \end{align}
    where 
    \begin{equation}
        z_{12} = \frac{z_1+z_2}{2}\;,
    \end{equation}
    and    
    \begin{equation}
D_{12} = D_\mathrm{A}(0,z_{12}),
    \end{equation}
    with the angular diameter distance $D_\mathrm{A}(z_1, z_2)$ between redshifts $z_1$ and $z_2$.
    The projected excess mass density $\Delta \Sigma$ is
    \begin{equation}
    \label{eq:definition Delta Sigma}
    \Delta \Sigma (\va*{\theta}, z_{\textrm{d}}) =
    \dfrac{\gamma_\textrm{t}(\va*{\theta})}{\bar{\Sigma}^{-1}_\textrm{crit}(z_\textrm{d})}\,,
    \end{equation}
    with the source-averaged inverse critical surface mass density $\bar{\Sigma}^{-1}_\mathrm{crit}$. Using the source redshift distribution $p(z_\textrm{s})$, $\bar{\Sigma}^{-1}_\mathrm{crit}$ is 
    \begin{equation}
    \label{eq:definition Sigma Crit inverse}
    \bar{\Sigma}^{-1}_\textrm{crit}(z_\textrm{d}) = \int_{z_\textrm{d}}^{\infty} \dd{z_\textrm{s}}\; p(z_\textrm{s})\,\frac{4\pi\, G}{c^2}\, \frac{D_\textrm{A}(z_\textrm{d}, z_\textrm{s})\, D_\textrm{A}(z_\textrm{d})}{D_\textrm{A}(z_\textrm{s})}\,,
    \end{equation}
    where $D_\textrm{A}(z) := D_\textrm{A}(0,z)$. This critical surface mass density is not the comoving critical surface mass density $\bar{\Sigma}_{\textrm{crit, com}}$, defined by  
    \begin{equation*}
        \label{eq:definition Sigma Crit inverse_com}
        \bar{\Sigma}^{-1}_\textrm{crit,com}(z_\textrm{d}) = \int_{z_\textrm{d}}^{\infty} \dd{z_\textrm{s}}\; p(z_\textrm{s})\,\frac{4\pi\, G}{c^2}\, \frac{D_\textrm{A}(z_\textrm{d}, z_\textrm{s})\, D_\textrm{A}(z_\textrm{d})}{(1+z_\mathrm{d})\, D_\textrm{A}(z_\textrm{s})}\,,
        \end{equation*}
    used in some weak lensing studies. Appendix C in \citet{Dvornik2018} discusses the implications of different definitions of the critical surface mass density.
        
    The correlation functions $\tilde{\mathcal{G}}_Z$ and $ \tilde{\mathcal{G}}_\textrm{phys}$ depend only on the lens-source distances $\vartheta_1$, $\vartheta_2$ and $r_1$, $r_2$, and the opening angle $\phi$ between the lens-source separations, because of the statistical isotropy of the matter and galaxy density fields.
    Therefore, we define
    \begin{equation}
    \tilde{\mathcal{G}}_Z(\vartheta_1, \vartheta_2, \phi):=\tilde{\mathcal{G}}_Z(\va*{\vartheta}_1, \va*{\vartheta}_2)\,,
    \end{equation}
    and
    \begin{equation}
    \tilde{\mathcal{G}}_\textrm{phys}(r_1, r_2, \phi):=\tilde{\mathcal{G}}_\textrm{phys}(\va*{r}_1, \va*{r}_2)\,.
    \end{equation}

    \subsection{Aperture statistics}
    \label{sec:theory:aperture statistics}
    The three-point correlation function $\tilde{\mathcal{G}}$ contains second- and third-order statistics. This can be seen by writing $\tilde{\mathcal{G}}$ as
    \begin{align}
    \label{eq:GtildeWithGGL}
    \tilde{\mathcal{G}}({\vartheta}_1, {\vartheta}_2, \phi) &= \expval{\kappa_\textrm{g}(\va*{\theta}+\va*{\vartheta}_1)\, \kappa_\textrm{g}(\va*{\theta}+\va*{\vartheta}_2)\, \gamma_\textrm{t}(\va*{\theta})}\\
    &\notag \quad + \expval{\kappa_\textrm{g}(\va*{\theta}+\va*{\vartheta}_1)\, \gamma_\textrm{t}(\va*{\theta})} + \expval{\kappa_\textrm{g}(\va*{\theta}+\va*{\vartheta}_2)\, \gamma_\textrm{t}(\va*{\theta})}\:,
    \end{align}
    where $\kappa_\textrm{g}(\va*{\vartheta})=N(\va*{\vartheta})/\bar{N}-1$ is the two-dimensional galaxy number density contrast. The second and third term in Eq.~\eqref{eq:GtildeWithGGL} are GGL statistics which correspond to the shear around individual lens galaxies. 
    
    When studying G3L we are interested in the excess shear around lens pairs, which is given only by the first term in Eq.~\eqref{eq:GtildeWithGGL}. Therefore, we converted $\tilde{\mathcal{G}}_Z$ and $\tilde{\mathcal{G}}_\textrm{phys}$ to the third-order aperture statistics $\expval{\mathcal{NN} M_\mathrm{ap}}$ and $\expval{\mathcal{NN} M_\mathrm{ap}}_\textrm{phys}$, 
    which only include the third-order statistics, as shown in \citetalias{Linke2020}. For this, we use a compensated filter function with an aperture scale $\theta$,
    \begin{equation}
    U_\theta(\vartheta)=\frac{1}{\theta^2}\, u\left(\frac{\vartheta}{\theta}\right)\, ,
    \end{equation}
    which fulfills
    \begin{equation}
    \int \dd{\vartheta}\; \vartheta\, U_\theta(\vartheta) = 0\;.
    \end{equation}
    With this filter function, the third-order aperture statistics are defined as
    \begin{align}
    \label{eq:definition NNMap}
    &\notag    \expval{\mathcal{NN} M_\mathrm{ap}}(\theta_1, \theta_2, \theta_3) \\
    =& \left[\int_{0}^{\infty} \dd{z_1} \, \int_{0}^{\infty} \dd{z_2}\; Z(\Delta z_{12})\, \bar{n}_1(z_1)\, \bar{n}_2(z_2)\right]^{-1}\\
    &\notag \times \int_{0}^{\infty} \dd{z_1}\, \int_{0}^{\infty} \dd{z_2}\, Z(\Delta z_{12}) \left[ \prod_{i=1}^{3} \int \dd[2]{\vartheta_i}\; \frac{1}{\theta_i^2}\, u\left(\frac{\vartheta_i}{\theta_i}\right)\right]\, \\
    &\notag\times \expval{n_1(\va*{\vartheta}_1, z_1)\, n_2(\va*{\vartheta}_2, z_2)\, \kappa(\va*{\vartheta}_3)}\, ,
    \end{align}
    and
    \begin{align}
    \label{eq:definition NNMapPhys}
    &\notag \expval{\mathcal{NN} M_\mathrm{ap}}_\textrm{phys}(r_1, r_2, r_3)\\ 
    =& \left[\int_{0}^{\infty} \dd{z_1} \, \int_{0}^{\infty} \dd{z_2}\; Z(\Delta z_{12})\, \bar{n}_1(z_1)\, \bar{n}_2(z_2)\right]^{-1}\\
    &\notag \times \int_{0}^{\infty} \dd{z_1}\, \int_{0}^{\infty} \dd{z_2}\, Z(\Delta z_{12})\, \left[\prod_{i=1}^{3} \int \dd[2]{x_i}\; \frac{1}{D_{12}^{-2}\,r^2_i}\,u\left(\frac{x_i}{r_i}\right)\right]\,\\
    &\notag \times  \expval{n_1(D_{12}^{-1}\,\va*{x}_1, z_1)\, n_2(D_{12}^{-1}\,\va*{x}_2, z_2)\, \Sigma(D_{12}^{-1}\,\va*{x}_3, z_{12}}\, , 
    \end{align}
    with the lensing convergence $\kappa(\va*{\theta})$ and the surface mass density $\Sigma$, given by
    \begin{equation}
    \Sigma(\theta, z) = \frac{\kappa(\theta)}{\bar{\Sigma}^{-1}_\textrm{crit}(z)}\;.
    \end{equation}  
    We use the exponential filter function
    \begin{equation}
    \label{eq:filter function}
    u(x)=\frac{1}{2\pi}\,\left(1-\frac{x^2}{2}\right)\, \exp(-\frac{x^2}{2})\;,
    \end{equation}
    for which the aperture statistics can be calculated from $\tilde{\mathcal{G}}_Z$ and $\tilde{\mathcal{G}}_\textrm{phys}$ with
    \begin{align}
    \label{eq:NNMap from Gtilde}
    &\expval{\mathcal{NN} M_\mathrm{ap}}(\theta_1, \theta_2, \theta_3) \\
    =&\notag \int_0^{\infty} \dd{\vartheta_1} \vartheta_1 \int_{0}^{\infty} \dd{\vartheta_2} \vartheta_2 \int_{0}^{2\pi} \dd{\phi} \; \tilde{\mathcal{G}}_Z(\vartheta_1, \vartheta_2, \phi)\,\\
    &\notag \times \mathcal{A}_{\mathcal{N}\mathcal{N}\mathcal{M}}(\vartheta_1, \vartheta_2, \phi \, | \, \theta_1, \theta_2, \theta_3)\,.
    \end{align}
    and
    \begin{align}
    \label{eq:NNMapPhys from GtildePhys}
    &\expval{\mathcal{NN} M_\mathrm{ap}}_\textrm{phys}(r_1, r_2, r_3) \\
    =&\notag \int_0^{\infty} \dd{x_1} x_1 \int_{0}^{\infty} \dd{x_2} x_2 \int_{0}^{2\pi} \dd{\phi} \; \tilde{\mathcal{G}}_\textrm{phys}(x_1, x_2, \phi)\, \\
    &\notag \times  \mathcal{A}_{\mathcal{N}\mathcal{N}\mathcal{M}}(x_1, x_2, \phi \, | \, r_1, r_2, r_3)\,.
    \end{align}
    The kernel function $\mathcal{A}_{\mathcal{N}\mathcal{N}\mathcal{M}}$ is defined in the appendix of \citet{Schneider2005}.
    
    \subsection{Galaxy bias}
    
    The aperture statistics can be used to constrain the galaxy bias, which is the relation between the galaxy number density contrast and the matter density contrast \citep[e.g][]{Schneider2005}. The simplest assumption for this bias is a linear deterministic relation,
    \begin{equation}
    \label{eq:linearBias}
    \kappa_\textrm{g}(\va*{\theta})=b\, \kappa(\va*{\theta})\,,
    \end{equation}
    where $b$ is a scale-independent bias factor \citep{Kaiser1984}. 
    A larger bias factor $b$ implies a higher galaxy number density for a given matter overdensity. For two galaxy populations with bias factors $b_1$ and $b_2$, this simple model predicts for the aperture statistics
    \begin{align}
    \label{eq:apertureStats and bias}
    &\expval{\mathcal{N}_1\,\mathcal{N}_2\,M_\textrm{ap}} \propto b_1\,b_2\,.        
    \end{align}
    From this follows, that
    \begin{equation}
    \label{eq:Biasratio}
    R:=\frac{\expval{\mathcal{N}_1\,\mathcal{N}_2\,M_\textrm{ap}} }{\sqrt{\expval{\mathcal{N}_1\,\mathcal{N}_1\,M_\textrm{ap}}\, \expval{\mathcal{N}_2\,\mathcal{N}_2\,M_\textrm{ap}} }}=    \frac{b_1\,b_2}{\sqrt{b_1^2\,b_2^2}} = 1\;.
    \end{equation}
    We measure $R$ in the observation and simulation to assess the assumption of linear deterministic bias. 
    

    \section{Methods}
    \label{sec:methods}
    \subsection{Estimating the three-point correlation function}
    \label{sec:methods:Gtilde}
    
    To measure $\tilde{\mathcal{G}}_{Z}$ and $\tilde{\mathcal{G}}_\textrm{phys}$, we used the estimators from \citetalias{Linke2020} for $N_\mathrm{s}$ source and $N_\mathrm{d}$ lens galaxies. These estimators measure the correlation functions by averaging the ellipticities of the source galaxies over all lens-lens-source galaxy triplets. For $\tilde{\mathcal{G}}_{Z}$, in the bin $B$ of $(\vartheta_1, \vartheta_2, \phi)$, the estimator is the real part of
    \begin{align}
    &\notag \tilde{\mathcal{G}}_{Z, \textrm{est}}(B) \\
    &=- \frac{\sum\limits_{i,j=1}^{N_\textrm{d}}\sum\limits_{k=1}^{N_\textrm{s}}w_k\, \epsilon_k\, \textrm{e}^{-\textrm{i}(\varphi_{ik} + \varphi_{jk})}\left[1 + \omega_Z(|\va*{\theta}_i - \va*{\theta}_j|)\right]Z(\Delta z_{ij})\, \Delta_{ijk}(B)}{\sum\limits_{i,j=1}^{N_\textrm{d}}\sum\limits_{k=1}^{N_\textrm{s}}\, w_k\,Z(\Delta z_{ij})\, \Delta_{ijk}(B)}\\
    &:=-\dfrac{\sum\limits_{ijk} \, w_k \, \epsilon_k \, \textrm{e}^{-\textrm{i}(\varphi_{ik}+\varphi_{jk})} \,\left[1+\omega_Z\left(|\va*{\theta}_i - \va*{\theta}_j|\right)\right]\, Z(\Delta z_{ij}) \, \Delta_{ijk}(B)}{\sum\limits_{ijk} w_k \,Z(\Delta z_{ij}) \,\Delta_{ijk}(B)}\;,
    \end{align}
    where $w_k$ is the weight of the source ellipticity $\epsilon_k$, $\omega_Z$ is the redshift-weighted angular two-point correlation function of the lens galaxies, and
    \begin{equation}
    \Delta_{ijk}(B) = \begin{cases}
    1 &\textrm{for } (|\va*{\theta}_k - \va*{\theta}_i|, |\va*{\theta}_k - \va*{\theta}_j|, \phi_{ijk})\in B\\
    0 &\textrm{otherwise}
    \end{cases}\,.
    \end{equation}
    The angles $\varphi_{ik}$ and $\varphi_{jk}$ are the polar angles of the lens-source separation vectors $\va*{\theta}_i-\va*{\theta}_k$ and $\va*{\theta}_j-\va*{\theta}_k$ (corresponding to $\varphi_1$ and $\varphi_2$ in Fig.~\ref{fig:G3L}) and $\phi_{ijk}=\varphi_{ik}-\varphi_{jk}$ is the opening angle between the lens-source separation vectors (corresponding to $\phi$ in Fig.~\ref{fig:G3L}). 
    
Source galaxies with more precise shape measurements receive a higher ellipticity weight $w_k$. The weight, therefore, increases the contribution of source galaxies with more exact shapes to the estimator. For the simulated shear data, we set the weights to $w_k=1$ for all sources.
    
    We obtained the lens two-point correlation function $\omega_Z$, with the estimator by \citet{Szapudi1998} which is 
    \begin{equation}
    \label{eq:szapudi&szalay}
    \omega_Z(\theta) = \frac{N_{\textrm{r}_1}\, N_{\textrm{r}_2}}{N_{\textrm{d}_1}\,N_{\textrm{d}_2}}\frac{{D_1D_2}_Z(\theta)}{{R_1R_2}_Z(\theta)} - \frac{N_{\textrm{r}_1}}{N_{\textrm{d}_1}}\frac{{D_1R_2}_Z(\theta)}{{R_1R_2}_Z(\theta)} - \frac{N_{\textrm{r}_2}}{N_{\textrm{d}_2}}\frac{{D_2R_1}_Z(\theta)}{{R_1R_2}_Z(\theta)} +1\;,
    \end{equation}
    for two different observed lens samples with $N_{\textrm{d}_1}$ and $N_{\textrm{d}_2}$ galaxies and two "random samples". These random samples contain $N_{\textrm{r}_1}$ and $N_{\textrm{r}_2}$ unclustered galaxies following the same selection function as the observed galaxies.
    
    The ${D_1D_2}_Z$, ${D_1R_2}_Z$, ${D_2R_1}_Z$,  and ${R_1R_2}_Z$ are the pair counts of observed and random galaxies.  For two equal lens samples and $DD = D_1D_2$, $DR = D_1R_2 = D_2R_1$, and $RR=R_1R_2$, the estimator in Eq.~\eqref{eq:szapudi&szalay} reduces to the usual Landy-Szalay estimator \citep{LandySzalay1993}
    \begin{equation}
    \label{eq:landy&szalay}
    \omega_Z(\theta) = \frac{N_{\textrm{r}}^2\, {DD}_Z(\theta)}{N_{\textrm{d}}^2\, {RR}_Z(\theta)} - 2\frac{N_{\textrm{r}}\, {DR}_Z(\theta)}{N_{\textrm{d}}\, {RR}_Z(\theta)}+1\,.
    \end{equation}
    Usually, pair counts are defined as the number of pairs within a certain angular separation. However, we used redshift-weighted pair counts to account for the stronger clustering of true lens pairs due to the redshift weighting function $Z$. They are defined for a bin centred at $\theta$ with bin size $\Delta \theta$ with the Heaviside step function $\Theta_\textrm{H}$ as
    \begin{align}
    \label{eq:ModifiedPaircountsDDRR}
    {D_1D_2}_Z(\theta) = \sum_{i=1}^{N_{\textrm{d}_1}} \sum_{j=1}^{N_{\textrm{d}_2}}&\Theta_\textrm{H}\left(\theta+{\Delta \theta}/{2} - |\va*{\theta}_i - \va*{\theta}_j|\right)\,\\
    &\notag\times \Theta_\textrm{H}\left(-\theta+{\Delta \theta}/{2} + |\va*{\theta}_i - \va*{\theta}_j|\right)\, Z(\Delta z_{ij})\; , \\
    {R_1R_2}_Z(\theta) = \sum_{i=1}^{N_{\textrm{r}_1}} \sum_{j=1}^{N_{\textrm{r}_2}}&\Theta_\textrm{H}\left(\theta+{\Delta \theta}/{2} - |\va*{\theta}_i - \va*{\theta}_j|\right)\\
    &\notag\times \Theta_\textrm{H}\left(-\theta+{\Delta \theta}/{2} + |\va*{\theta}_i - \va*{\theta}_j|\right)\, Z(\Delta z_{ij})\; ,
    \end{align}
    and
    \begin{align}
    \label{eq:ModifiedPaircountsDR}
    {D_aR_b}_Z(\theta) = \sum_{i=1}^{N_{\textrm{d}_a}} \sum_{j=1}^{N_{\textrm{r}_a}}& \Theta_\textrm{H}\left(\theta+{\Delta \theta}/{2} - |\va*{\theta}_i - \va*{\theta}_j|\right)\\
    &\notag\times \Theta_\textrm{H}\left(-\theta+{\Delta \theta}/{2} + |\va*{\theta}_i - \va*{\theta}_j|\right)\, Z(\Delta z_{ij})\;,
    \end{align}
    with $a,b \in {1,2}$.
    
    Following \citetalias{Linke2020}, we chose a Gaussian weighting function,
    \begin{equation}
    Z(\Delta z_{12}) = \exp(-\frac{\Delta z_{12}^2}{2\sigma_z^2})\, ,
    \end{equation}
    with width $\sigma_z = 0.01$. This width is larger than typical galaxy correlation lengths and redshifts induced by the peculiar motion of galaxies. Accordingly, true lens pairs are not affected by the redshift weighting, while chance pairs are down-weighted. Choosing a different $\sigma_z$ influences the magnitude of the measured aperture statistics as well as the S/N of the measurement. Nonetheless, as long as the same width is chosen for the observation and the simulation, their G3L signals can be compared.
    
    The estimator for $\tilde{\mathcal{G}}_\textrm{phys}$ for the bin $B$ of $(r_1, r_2, \phi)$ is
    \begin{align}
    \label{eq:3ptcorrelation_estimator_physical}
    &\tilde{\mathcal{G}}_\textrm{est,phys}(B)=\\
    &\notag - \dfrac{\sum\limits_{ijk} w_k \, \epsilon_k \, \textrm{e}^{-\textrm{i}(\varphi_{ik}+\varphi_{jk})}\left[1 + \omega_Z\left(|\va*{\theta}_i - \va*{\theta}_j|\right)\right]  Z(\Delta z_{ij})\, \bar{\Sigma}_{\textrm{crit}}^{-1}(z_{ij})\, \Delta_{ijk}^{\textrm{ph}}(B)}{\sum\limits_{ijk} w_k \, \bar{\Sigma}_{\textrm{crit}}^{-2}(z_{ij})\,  Z(\Delta z_{ij})\, \Delta_{ijk}^{\textrm{ph}}(B)}\;,
    \end{align}
    with 
    \begin{align}
    \label{eq:trianglePhys}
    &\Delta_{ijk}^{\textrm{ph}}(B) = 
    \begin{cases}
    1 &\textrm{for}\left(D_{ij}\,|\va*{\theta}_k-\va*{\theta}_i|,D_{ij}\,|\va*{\theta}_k-\va*{\theta}_j|,\phi_{ijk}\right) \in B \\
    0 &\textrm{otherwise}
    \end{cases}\,.
    \end{align}    
    We measured $\tilde{\mathcal{G}}_{Z}$ and $\tilde{\mathcal{G}}_\textrm{phys}$ initially for $128\times128\times128$ bins, which were linearly spaced along $\phi$ and logarithmically spaced along $\vartheta_{1,2}$ and $r_{1,2}$. For $\tilde{\mathcal{G}}_{Z}$, the $\vartheta_{1,2}$ are between $\ang[astroang]{;0.15;}$ and $\ang[astroang]{;200;}$ for the observed and between $\ang[astroang]{;0.15;}$ and $\ang{;320;}$ for the simulated data. For $\tilde{\mathcal{G}}_\textrm{phys}$, we chose $r_{1,2}$ between $0.02\,\textrm{Mpc}$ and $40\,\textrm{Mpc}$. We then applied the adaptive binning scheme of \citetalias{Linke2020}, by which the parameter space was tessellated to remove bins for which no galaxy triplet is in the data.
    
    The correlation function was measured individually for 24 tiles of the observational data of size $\ang{2.5}\times\ang{3}$ and 64 fields-of-view of the MR of size $\ang{4}\times\ang{4}$, leading to estimates $\tilde{\mathcal{G}}_\textrm{est}^i$ and $\tilde{\mathcal{G}}_\textrm{est, ph}^i$  for each tile and field-of-view, respectively. The division into small patches allowed us to project the observational measurements to Cartesian coordinates and to estimate the uncertainty of the measurement with jackknife resampling. For each data set, the individual estimates were combined to form the total correlation functions with
    \begin{equation}
    \tilde{\mathcal{G}}_\textrm{est}(B)=\frac{\sum^N_{i=1} \tilde{\mathcal{G}}_\textrm{est}^i(B)\,W^i(B)}{\sum^N_{i=1}W^i(B)}\; ,
    \end{equation}
    where
    \begin{equation}
    W^i(B)=\sum\limits_{ijk} w_k \,Z(\Delta z_{ij})\, \,\Delta_{ijk}(B)\;,
    \end{equation}
    and
    \begin{equation}
    \tilde{\mathcal{G}}_\textrm{est, ph}(B)=\frac{\sum_{i=1}^{N} \tilde{\mathcal{G}}_\textrm{est, ph}^i(B)\,W_\textrm{ph}^i(B)}{\sum^N_{i=1}W_\textrm{ph}^i(B)}\,,
    \end{equation}
    where
    \begin{equation}
    W^i_\textrm{ph}(B)=\sum\limits_{ijk} w_k \, \bar{\Sigma}_{\textrm{crit}}^{-2}(z_{ij})\,  Z(\Delta z_{ij})\, \Delta_{ijk}^{\textrm{ph}}(B)\;.
    \end{equation}
    
    \subsection{Computing aperture statistics}
    
    To compute $\expval{\mathcal{NN} M_\mathrm{ap}}$ and $\expval{\mathcal{NN} M_\mathrm{ap}}_\textrm{phys}$, 
    we integrated over $\tilde{\mathcal{G}}_Z$ using Eq.~\eqref{eq:NNMap from Gtilde} and \eqref{eq:NNMapPhys from GtildePhys}. 
    We numerically approximated the integrals by summing over all $N_\textrm{bins}$ of $\tilde{\mathcal{G}}_Z$ after tessellation with
    \begin{align}
    &\expval{\mathcal{NN} M_\textrm{ap}}(\theta_1, \theta_2, \theta_3)\\
    &\notag= \sum_{i=1}^{N_\textrm{bin}} V(b_i)\, \tilde{\mathcal{G}}_{Z,\textrm{est}}(b_i) \, \mathcal{A}_{NNM}(b_i\,|\,\theta_1, \theta_2, \theta_3)\, ,
    \end{align}
    and
    \begin{align}
    &\expval{\mathcal{NN} M_\textrm{ap}}_\textrm{phys}(r_1, r_2, r_3)\\
    &\notag =\sum_{i=1}^{N_\textrm{bin}} V(b_i)\, \tilde{\mathcal{G}}_{\textrm{est, phys}}(b_i) \, \mathcal{A}_{NNM}(b_i\,|\,r_1, r_2, r_3)\,.
    \end{align}
    Here, $b_i$ is the $i$th bin, which has size $V(b_i)$, and $\mathcal{A}_\mathcal{NNM}$ is the kernel function evaluated at the tessellation seed of bin $b_i$.
    
    The aperture statistics were measured only for equal scale radii $\theta_1=\theta_2=\theta_3$, so we abbreviate
    \begin{align}
    &\expval{\mathcal{NN} M_\mathrm{ap}}(\theta, \theta, \theta) = \expval{\mathcal{NN} M_\mathrm{ap}}(\theta)\;,\\
    &\expval{\mathcal{NN} M_\mathrm{ap}}_\textrm{phys}(r,r,r) = \expval{\mathcal{NN} M_\mathrm{ap}}_\textrm{phys}(r)\,.
    \end{align}    
    We estimated the statistical uncertainty of the aperture statistics in the observational data with jackknife resampling. For this we assumed that the 24 tiles are statistically independent. Although this assumption is not correct for noise due to sample variance, we expect our noise to be dominated by shape noise which is independent for each tile. In the jackknife resampling, we combined the $\tilde{\mathcal{G}}^i_Z$ of all $N$ tiles to the total $\tilde{\mathcal{G}}_Z$, and also create $N$ jackknife samples, for which all but one tile are combined. The aperture statistics $\expval{\mathcal{NN} M_\mathrm{ap}}(\theta)$ are calculated for the total $\tilde{\mathcal{G}}_Z$, as well as for each of the $N$ jackknife samples to get $N$ $\expval{\mathcal{NN} M_\mathrm{ap}}_k(\theta)$. The covariance matrix of $\expval{\mathcal{NN} M_\mathrm{ap}}(\theta)$ was then estimated with
    \begin{align}
    \label{eq:covariancematrix}
    C_{ij}= \frac{N}{N-1}\sum_{k=1}^{N}& \left[\expval{\mathcal{NN} M_\mathrm{ap}}_k(\theta_i)-\overline{\expval{\mathcal{NN} M_\mathrm{ap}}_k}(\theta_i)\right]\\
    &\notag \times \left[\expval{\mathcal{NN} M_\mathrm{ap}}_k(\theta_j)-\overline{\expval{\mathcal{NN} M_\mathrm{ap}}_k}(\theta_j)\right]\;,
    \end{align}
    where $\overline{\expval{\mathcal{NN} M_\mathrm{ap}}_k}(\theta_i)$ is the average of all $\expval{\mathcal{NN} M_\mathrm{ap}}_k(\theta_i)$. 
    
    As discussed by \citet{Hartlap2007} and \citet{Anderson2003}, the inverse of this estimate of the covariance matrix is not an unbiased estimate of the inverse covariance matrix. Following their 
    suggestion, we instead estimated the inverse covariance matrix with
    \begin{equation}
    \label{eq:inverseCovariancematrix}
    C^{-1}_{ij} = \frac{N}{N-p-1} \left(C_{ij}\right)^{-1}\;,
    \end{equation}
    where $p$ is the number of data points. This gives an unbiased estimate of the inverse covariance matrix if the realisations are statistically independent and have Gaussian errors.
    
    With this estimate of the inverse covariance matrix, we calculated the S/N of our observational measurement with
    \begin{equation}
    \label{eq:S/N}
    \textrm{S/N} = \left[\sum_{i,j=1}^{p} \expval{\mathcal{NN} M_\mathrm{ap}}(\theta_i)\, C^{-1}_{ij}\, \expval{\mathcal{NN} M_\mathrm{ap}}(\theta_j)\right]^{1/2}\;.
    \end{equation}
    We also use $C^{-1}_{ij}$ to perform a $\chi^2$-test, evaluating the agreement of the observational measurement $\expval{\mathcal{NN} M_\mathrm{ap}}_{\rm obs}$ with the SAMs prediction $\expval{\mathcal{NN} M_\mathrm{ap}}_{\rm sim}$. For this, we calculate the reduced $\chi^2_\mathrm{redu}$ as 
    \begin{align}
    \chi^2_\textrm{redu} &= \frac{1}{p}\,\sum_{i,j=1}^{p} \left(\expval{\mathcal{NN} M_\mathrm{ap}}_{\rm obs}(\theta_i) - \expval{\mathcal{NN} M_\mathrm{ap}}_{\rm sim}(\theta_i)\right)\\
    &\notag \quad\quad\quad \times C^{-1}_{ij} \left(\expval{\mathcal{NN} M_\mathrm{ap}}_{\rm obs}(\theta_j) - \expval{\mathcal{NN} M_\mathrm{ap}}_{\rm sim}(\theta_j)\right)\;.
    \end{align}
    
    Aside from $\expval{\mathcal{NN} M_\mathrm{ap}}_\textrm{phys}$, we also estimated the so-called B-mode $\expval{\mathcal{NN} M_{\perp}}$, given by applying Eq.~\eqref{eq:NNMapPhys from GtildePhys} not on $\tilde{\mathcal{G}}_\textrm{phys}$ but on
    \begin{align}
    \label{eq:GtildeCross}
    &\tilde{\mathcal{G}}_\perp(B)\\
    &\notag=-\dfrac{\sum\limits_{ijk} \, w_k \, {\epsilon}^*_k \, \textrm{e}^{\textrm{i}(\varphi_{ik}+\varphi_{jk})}\left[1+\omega_Z\left(|\va*{\theta}_i - \va*{\theta}_j|\right)\right]\, Z(\Delta z_{ij}) \, \bar{\Sigma}^{-1}_\textrm{crit}\,\Delta^\textrm{ph}_{ijk}(B)}{\sum\limits_{ijk} w_k\,\bar{\Sigma}^{-2}_\textrm{crit} \,Z(\Delta z_{ij}) \,\Delta^\textrm{ph}_{ijk}(B)}\,,
    \end{align}
    where ${\epsilon}^*_k$ is the complex conjugate of the galaxies ellipticity. If there are no dominating systematic effects inducing a parity violation, the B-mode has to vanish \citep{Schneider2003}. We therefore tested for such systematics by estimating $\expval{\mathcal{NN} M_{\perp}}$. 
    
    \section{Data}
    \label{sec:data}
    
    \subsection{Observational data}
    \label{sec:data:observations}
    Our observational data is the overlap of KiDS, VIKING, and GAMA (KV450 $\times$ GAMA). This overlap encompasses approximately 180 $\textrm{deg}^2$, divided into the three patches G9, G12 and G15, each with dimensions of $12 \times 5 \, \textrm{deg}^2$. 
    
    VIKING \citep{Edge2013, Venemans2015} is a photometric survey in five near-infrared bands, conducted at the VISTA telescope in Paranal, Chile and covering approximately $1350\, \textrm{deg}^2$. It covers the same area as KiDS \citep{Kuijken2015, deJong2015}, an optical photometric survey conducted with the OmegaCAM at the VLT Survey Telescope. The data of KiDS and VIKING were combined to form the KV450 data set, described in detail in \citet{Wright2019}, which we use in the following. KV450 has the same footprint as the third data release of KiDS \citep{deJong2017} and was processed by the same data reduction pipelines, described in detail in \citet{Hildebrandt2017}. Data are processed by THELI \citep{Erben2005, Schirmer2013} and Astro-WISE \citep{deJong2015}. Shears are measured with \emph{lens}fit \citep{Miller2013, Kannawadi2019}. Photometric redshifts are obtained from PSF-matched photometry \citep{Wright2019} and calibrated using external overlapping spectroscopic surveys \citep{Hildebrandt2020}.
    
    We use the galaxies observed by KV450 with photometric redshift between $0.5$ and $1.2$ as source galaxies. Galaxies with a photometric redshift less than $0.5$ are excluded because most of them are in front of our lens galaxies and therefore dilute and bias the lensing signal. The averaged inverse critical surface mass density $\bar{\Sigma}^{-1}_\mathrm{crit}$ is calculated as described in Sect.~\ref{sec:methods} by using the weighted direct calibration redshift distributions (DIR distributions) of the KV450 galaxies as the source distribution. These DIR distributions were obtained with in-depth spectroscopic surveys overlapping with KiDS and VIKING. The spectroscopic redshift distributions from these surveys were weighted according to the photometric data in KV450 to estimate the redshift distribution of KV450 galaxies. Details of this procedure are given in \citet{Hildebrandt2017, Hildebrandt2020}. We neglect the uncertainties on the redshift distribution and the multiplicative bias of the shear estimate. However, as these uncertainties are small, we do not expect them to impact our conclusions.

    GAMA \citep{Driver2009, Driver2011, Liske2015} is a spectroscopic survey carried out at the Anglo Australian Telescope with the AAOmega spectrograph. We use the data management unit (DMU) \verb|distanceFramesv14|, which contains positions and spectroscopic redshifts $z$ of galaxies with a Petrosian observer-frame $r$-band magnitude brighter than 19.8 mag. The spectroscopic redshifts were flow-corrected to account for the proper motion of the Milky Way using the model by \citet{Tonry2000} according to the procedure in \citet{Baldry2012}. We include all galaxies with a spectroscopic redshift lower than 0.5 and redshift quality flag \verb|N_Q|$\ge$3. For the calculation of the angular two-point correlation function of lenses, we use randoms from the DMU \verb|randomsv02| \citep{Farrow2015}, which incorporates the galaxy selection function of GAMA while maintaining an unclustered galaxy distribution.
    
    From the GAMA galaxies, we select lens samples according to their colour and stellar mass. Restframe photometry and stellar masses were obtained from the DMU \verb|stellarMassesLambdarv20|. An overview of our samples is given in Table~\ref{tab:sub-samples}.
      
    We select a `red' and `blue' lens sample, defined according to the galaxies' rest-frame $(g-r)_0$ colour. We use the colour cut by \citet{Farrow2015}, according to which a galaxy is red if its rest-frame colour $(g-r)_0$ and its absolute Petrosian magnitude $M_r$ in the $r$-band fulfil
    \begin{equation}
    (g-r)_0 + 0.03\, (M_r - 5 \log_{10}h+ 20.6) > 0.6135\,.
    \end{equation}
    Otherwise, the galaxy is considered blue. This colour cut is chosen to yield approximately equal numbers of red and blue galaxies ($93\,524$ red and $93\,702$ blue galaxies). Using a hard colour cut does not automatically produce two physically distinct galaxy populations \citep{Taylor2015}. However, as we apply the same cuts in the observational and simulated data, we expect to obtain comparable ``red'' and ``blue'' galaxy samples.
    
    Absolute magnitudes and rest-frame colours of the GAMA galaxies were obtained by \citet{Wright2016} using matched aperture photometry and the LAMBDAR code. These magnitudes were aperture corrected, using
    \begin{equation}
    M_{r, \textrm{tot}} = M_{r, \textrm{meas}} - 2.5 \log_{10}f + 5\log_{10}h\,,
    \end{equation}
    where $f$ is the flux scale, which is the ratio between the measured $r$-band flux and the total $r$-band flux inferred from fitting a S\'ersic-profile to the galaxies photometry.
    
    We define five stellar mass bins with the same cuts as \citet{Farrow2015}, with $M^*$ between $10^{8.5}\,h^{-2}\,\mathrm{M}_{\odot}$ and $10^{11.5}\,h^{-2}\,\mathrm{M}_{\odot}$. The stellar masses of GAMA galaxies were obtained by \citet{Wright2017}, assuming the initial mass function by \citet{Chabrier2003}, stellar population synthesis according to \citet{Bruzual2003}, and dust extinction according to \citet{Calzetti2000}. 
    
    \begin{table*}
        \caption{Selection criteria for lens samples and number density $N$ of selected galaxies per sample.}
        \label{tab:sub-samples}
        \centering
        \begin{tabular}{ccccc}
            \hline\hline
            Sample & {Selection Criterion} & $N$ (GAMA) [$\mathrm{arcmin}^{-2}$] &  $N$ (\citetalias{Henriques2015}) [$\mathrm{arcmin}^{-2}$] & $N$ (\citetalias{Lagos2012}) [$\mathrm{arcmin}^{-2}$]\\
            \hline
            m1 & $8.5<\log_{10}(M^*/\mathrm{M}_{\odot}\, h^{-2})\le9.5$ & $0.037$ & $0.040$ & $0.059$\\
            m2 & $9.5<\log_{10}(M^*/\mathrm{M}_{\odot}\, h^{-2})\le10$ & $0.058$ & $0.059$ & $0.064$\\
            m3 & $10<\log_{10}(M^*/\mathrm{M}_{\odot}\, h^{-2})\le10.5$ & $0.099$ & $0.096$ & $0.095$\\
            m4 & $10.5<\log_{10}(M^*/\mathrm{M}_{\odot}\, h^{-2})\le11$ & $0.080$ & $0.076$ & $0.058$\\
            m5 & $11<\log_{10}(M^*/\mathrm{M}_{\odot}\, h^{-2})\le11.5$ & $0.014$ & $0.011$ & $0.009$\\
            \hline
            red & $(g-r)_0 + 0.03\, (M_r - 5 \log_{10}h+ 20.6) > 0.6135$  & 0.143 & 0.140 & 0.152\\
            blue & $(g-r)_0 + 0.03\, (M_r - 5 \log_{10}h+ 20.6) \le 0.6135$ & 0.144 & 0.142 & 0.139\\
            \hline
        \end{tabular}
    \tablefoot{Lenses are selected either according to their stellar mass $M^*$ or to their rest-frame $(g-r)_0$ colour and absolute $r$-band magnitude $M_r$ and need to have $r<19.8\,\mathrm{mag}$.}
    \end{table*}
    
    The estimator for $\tilde{\mathcal{G}}_Z$ and $\tilde{\mathcal{G}}_\textrm{phys}$ are defined in terms of Cartesian coordinates. Therefore, we project the right ascension $\alpha$ and the declination $\delta$ of the galaxies onto a tangential plane on the sky. For this, we divide the source and the lens galaxy catalogues into 24 tiles with a size of $2.5 \times 3 \, \textrm{deg}^2$, which are also used for the jackknife resampling. We use the tile centres $(\alpha_0, \delta_0)$ as projection points and find the Cartesian coordinates $(x,y)$ with the orthographic projection
    \begin{align}
    &x = \cos(\delta) \sin(\alpha - \alpha_0) \,,\\
    &y = \cos(\delta_0)\sin(\delta) - \sin(\delta_0)\cos(\delta)\cos(\alpha - \alpha_0)\,.
    \end{align}
    
    \subsection{Simulated data}
    We compare the results for the aperture statistics in KV450 $\times$ GAMA to measurements in the MR with two different SAMs. 
    
    The MR \citep{Springel2005} is a dark-matter-only cosmological $N$-body-simulation. It traces the evolution of $2160^3$ dark matter particles of mass $m=8.6\times 10^8\,h^{-1}\,\mathrm{M}_{\odot}$  from redshift $z=127$ to today in a cubic region with co-moving side length $500\, h^{-1}\,\textrm{Mpc}$.
    
    Maps of the gravitational shear $\gamma$, caused by the matter distribution in the MR, are created with the multiple-lens-plane ray-tracing algorithm by \citet{Hilbert2009}. With this algorithm, we obtain 64 maps of $\gamma$ on a regular mesh with $4096^2$ pixels, corresponding to $4 \times 4\, \textrm{deg}^2$ on a set of redshift planes. For each field-of-view, we combine the shear on nine redshift planes between $z=1.2$ and $z=0.5$ by averaging it, weighted according to the redshift distribution of the KV450 source galaxies. From this, we obtain shear maps, which have the same source galaxy distribution as the observational data. We use the DIR redshift distribution, whose creation we described in Sect.~\ref{sec:data:observations}, and do not add any shape noise to the shear.
    
    We obtain simulated lens galaxies from two SAMs implemented in the MR, the SAM by \citetalias{Henriques2015} and the SAM by \citetalias{Lagos2012}. The \citetalias{Henriques2015} SAM assumes the same initial mass function by \citet{Chabrier2003} as the observations, but a different stellar population model, that is, the one by \citet{Maraston2005}. The \citetalias{Lagos2012} SAM uses the initial mass function by \citet{Kennicutt1983} and the stellar population model of \citet{Bruzual2003}. While the magnitudes of the \citetalias{Henriques2015} SAM are given in AB-magnitudes, the magnitudes of the \citetalias{Lagos2012} SAM are originally in the Vega magnitude system. We convert the magnitudes to the AB-system with the conversion suggested by \citet{Blanton2007}, 
    \begin{align}
    g_\textrm{AB} &= g_\textrm{Vega} - 0.08\;,\\
    r_\textrm{AB} &= r_\textrm{Vega} + 0.16\;.
    \end{align}
    
\begin{figure}
        \resizebox{\hsize}{!}{\includegraphics[width=\linewidth, trim=0.2cm 0 1.4cm 0, clip]{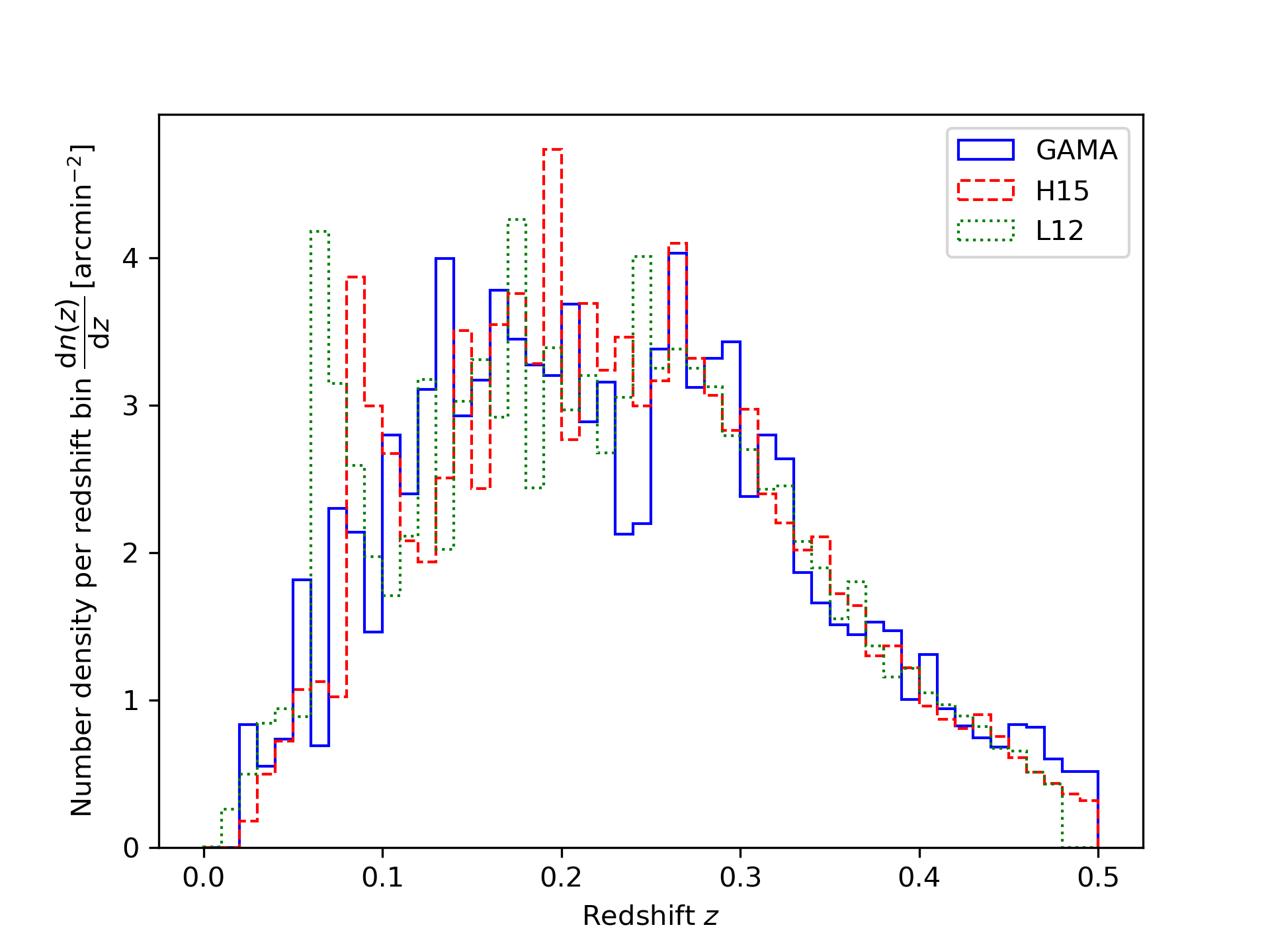}}
        \caption{Number density per redshift bin of GAMA (solid blue) \citetalias{Henriques2015} galaxies (dashed red), and \citetalias{Lagos2012} galaxies (dotted green) for the limiting magnitude of $r<19.8$. The bin size is $\Delta z=0.01$.}
        \label{fig:dn_dz}
    \end{figure}

   The lens galaxies are selected in the same way as the lenses in GAMA. We use all galaxies with redshifts less than $0.5$ and brighter than $r=19.8\,\textrm{mag}$, which is the limiting magnitude of GAMA. With this criterion, we aim to mimic the selection function of GAMA galaxies and expect to obtain samples of similar lenses as in the observation. Systematic errors in the galaxy fluxes, for example, due to the dust modelling of either GAMA or the SAM galaxies, could invalidate this expectation, as different galaxies would be sampled. However, as shown in Fig.~\ref{fig:dn_dz}, the redshift distribution of selected simulated and observed lens galaxies agree well. This likely would not be the case if there were fundamental differences in the selection function for simulated and observed galaxies. The number density of simulated lenses  $0.282\, \textrm{arcmin}^{-2}$ for the \citetalias{Henriques2015} SAM and $0.291\,\textrm{arcmin}^{-2}$ for the \citetalias{Lagos2012} SAM, which are both close to the GAMA number density of $0.287\, \textrm{arcmin}^{-2}$. Consequently, we expect the lens samples in the simulated and observational data to be comparable.
       
\begin{figure}
    \resizebox{\hsize}{!}{\includegraphics[width=\linewidth, trim=0.2cm 0 1.4cm 0, clip]{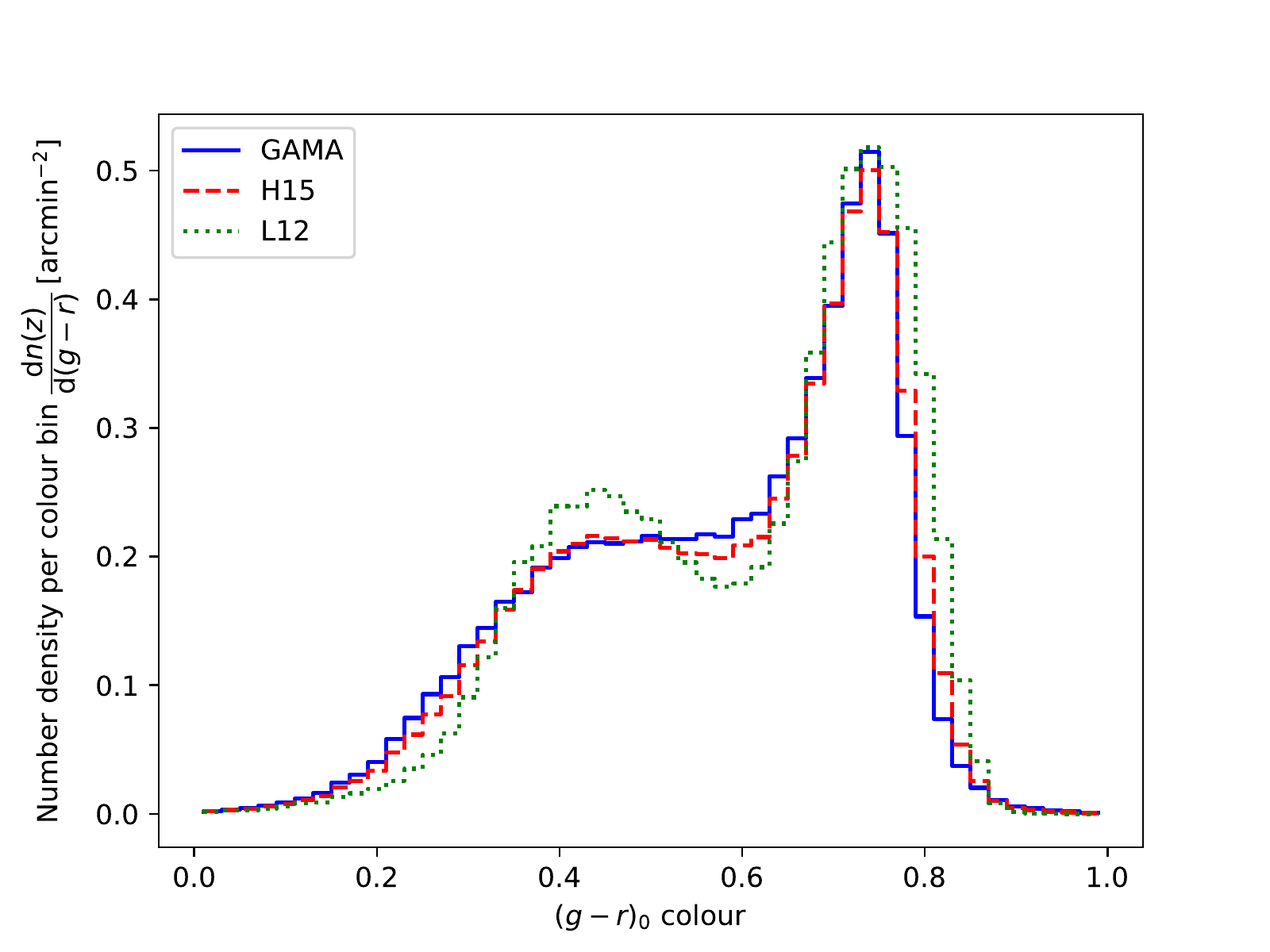}}
    \caption{Number density per colour bin of GAMA (solid blue) \citetalias{Henriques2015} (dashed red), and \citetalias{Lagos2012} galaxies (dotted green) for the limiting magnitude of $r<19.8$. The bin size is $\Delta (g-r)_0=0.01$.}
    \label{fig:colourdist}
\end{figure}

\begin{figure}
    \resizebox{\hsize}{!}{\includegraphics[width=\linewidth, trim=0 0 1.4cm 0, clip]{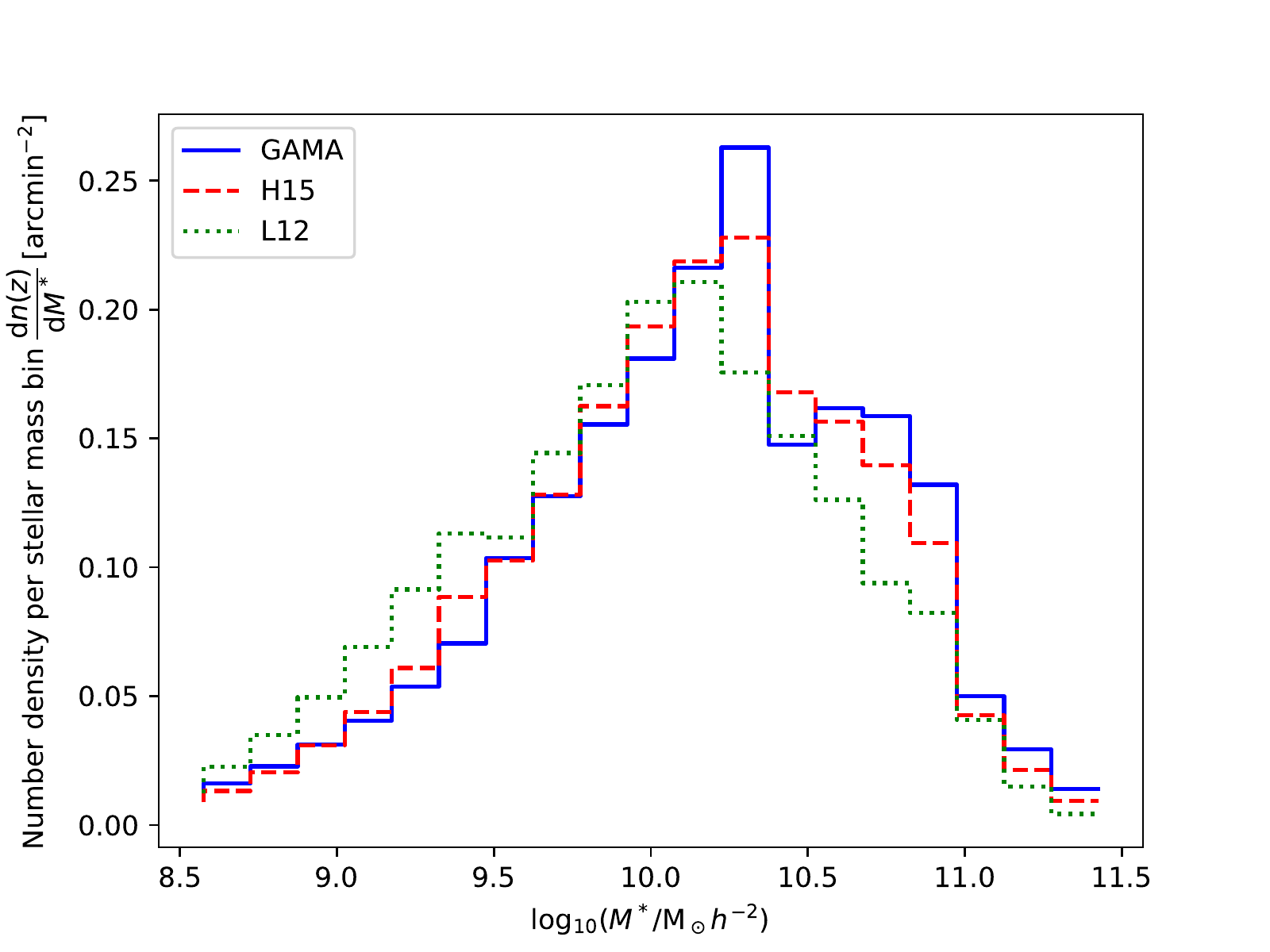}}
    \caption{Number density per stellar mass bin of GAMA (solid blue) \citetalias{Henriques2015} (dashed red), and \citetalias{Lagos2012} galaxies (dotted green) for the limiting magnitude of $r<19.8$. The bin size is $\Delta \log(M^*/M_\odot h^{-2})=0.15$.}
    \label{fig:massdist}
\end{figure}

    We split the simulated lens galaxies into colour and stellar-mass samples by applying the same cuts as to the GAMA galaxies (Table~\ref{tab:sub-samples}). Figures~\ref{fig:colourdist} and ~\ref{fig:massdist} show the colour- and stellar mass distribution of observed and simulated galaxies. The colour distributions of GAMA and SAM galaxies have similar modes. However, the blue mode of the \citetalias{Lagos2012} SAM is more concentrated. The \citetalias{Henriques2015} SAM also predicts stellar mass distributions similar to the observation, while the \citetalias{Lagos2012} SAM predicts more galaxies with stellar masses below $9.5\times 10^{10} \mathrm{M}_\odot$ and fewer galaxies with stellar masses above $11\times 10^{10} \mathrm{M}_\odot$.
    
    \section{Results}
    \label{sec:results}
    
    In this section, we present our results for the physical aperture statistics $\expval{\mathcal{NN} M_\mathrm{ap}}_\mathrm{phys}$, defined in Eq.~\eqref{eq:definition NNMapPhys}. The measured angular aperture statistics $\expval{\mathcal{NN} M_\mathrm{ap}}$, which exhibit similar trends, are given in Appendix~\ref{app:NNMap}.
    
    \begin{figure}
        \includegraphics[width=\linewidth, trim=0.2cm 0 1.4cm 0, clip]{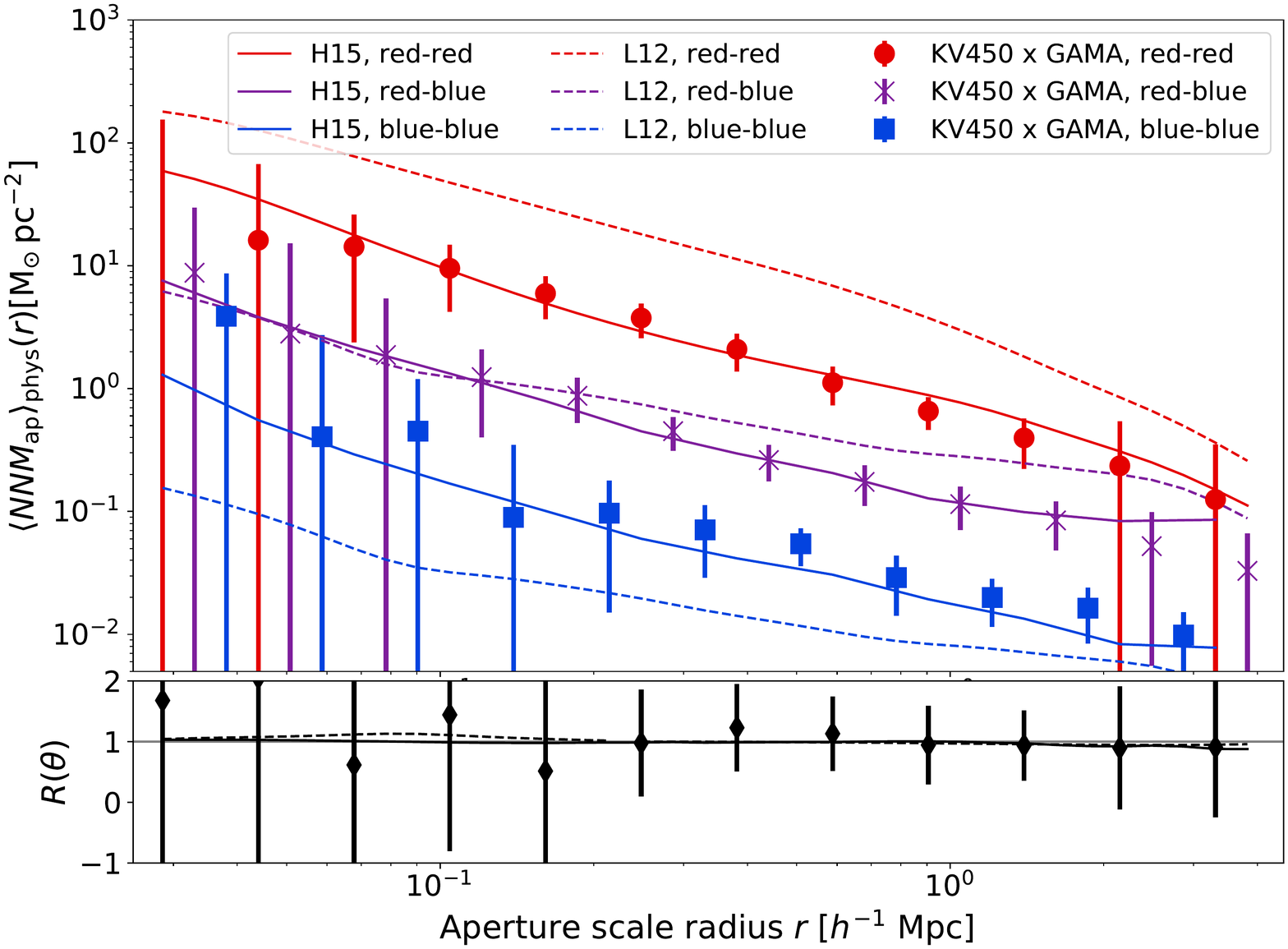}
        \caption{\emph{Upper panel:} Physical aperture statistics for colour-selected lens samples of the \citetalias{Henriques2015} galaxies (solid lines), \citetalias{Lagos2012} galaxies (dashed lines) and KV450 $\times$ GAMA (points). The signal is shown for red-red lens pairs (red lines and filled circles), red-blue lens pairs (purple lines and crosses), and blue-blue lens pairs (blue lines and squares). Error bars on the observational measurements are the standard deviation from jackknifing. \emph{Lower panel:} Ratio statistics $R$ as given by Eq.~\eqref{eq:Biasratio} for the red and blue lens samples of KV450 $\times$ GAMA (points), the \citetalias{Henriques2015} SAM (solid line) and the \citetalias{Lagos2012} SAM (dashed line). }
        \label{fig:NNMap_colours}
    \end{figure}
    
    The upper plot of Fig.~\ref{fig:NNMap_colours} presents $\expval{\mathcal{NN} M_\mathrm{ap}}_\mathrm{phys}$ for red-red, red-blue, and blue-blue lens pairs. For the observed and both simulated data sets, the signal for red-red lens pairs is larger than for red-blue and blue-blue lens pairs. Consequently, the linear deterministic bias model of Eq.~\eqref{eq:linearBias} suggests that the bias factor $b_\mathrm{red}$ of red galaxies is larger than the bias factor $b_\mathrm{blue}$ of blue galaxies. 
    
    The linear deterministic bias model predicts that the aperture statistics for mixed red-blue lens pairs are the geometric mean of the aperture statistics for red-red and blue-blue lens pairs (see Eq.~\ref{eq:Biasratio}). To test this prediction, we show $R$ in the lower plot of Fig.~\ref{fig:NNMap_colours}. For the observed galaxies, $R$ is consistent with unity, supporting the linear deterministic bias model. However, for scales below $0.2\,h^{-1}\,\mathrm{Mpc}$, the noise of the observed $R$ is more than three times larger than $R$ itself, which inhibits any meaningful deductions on the bias model at small scales.  For the \citetalias{Henriques2015} model, the prediction by the linear bias model is fulfilled , while for the \citetalias{Lagos2012} model $R$ is slightly larger than unity at scales below $0.2\,h^{-1}\,\mathrm{Mpc}$. 
    
    \begin{table}
        \caption{$\chi^2_\mathrm{redu}$ of $\expval{\mathcal{NN} M_\mathrm{ap}}_\textrm{phys}$ for the \citetalias{Henriques2015} and \citetalias{Lagos2012} SAMs.}
        \label{tab:chi2}
        \centering
        \begin{tabular}{ccccc}
            \hline\hline
            lens pairs  & $\chi^2_\textrm{redu}$ for \citetalias{Henriques2015} & $\chi^2_\textrm{redu}$ for \citetalias{Lagos2012}\\
            \hline
            red -- red & $1.08$ & \textbf{55.12} \\
            red -- blue & $0.95$  & \textbf{3.13}\\
            blue -- blue & $1.10$ & \textbf{2.19}\\
            \hline
            m1 -- m1& 1.44& \textbf{32.72}\\
            m1 -- m2& 1.75& \textbf{42.96}\\
            m1 -- m3& 1.54& \textbf{46.83}\\
            m1 -- m4& \textbf{2.84}& \textbf{45.33}\\
            m1 -- m5& 1.75& \textbf{64.22}\\
            m2 -- m2& 1.58& \textbf{16.04}\\
            m2 -- m3& 0.80& \textbf{17.11}\\
            m2 -- m4& 0.97& \textbf{10.04}\\
            m2 -- m5& 0.85& \textbf{47.10}\\
            m3 -- m3& 1.31& \textbf{51.21}\\
            m3 -- m4& 1.17& \textbf{41.01}\\
            m3 -- m5& 1.10& \textbf{8.60}\\
            m4 -- m4& 1.62& \textbf{2.56}\\
            m4 -- m5& 0.97& \textbf{8.54}\\
            m5 -- m5& 0.73& \textbf{6.93}\\
            \hline
        \end{tabular}
        \tablefoot{Samples are selected according to Table \ref{tab:sub-samples}. Bold values indicate a tension at the 95\% CL.}
    \end{table}
    
    The SAMs give different predictions for the aperture statistics. While the $\expval{\mathcal{NN} M_\mathrm{ap}}_\textrm{phys}$ of the \citetalias{Henriques2015} SAM agrees well with the observations, the signals for red-red and blue-blue lens pairs of the \citetalias{Lagos2012} model differ markedly. The \citetalias{Lagos2012} SAM predicts much larger aperture statistics for red-red pairs than the observation and significantly smaller aperture statistics for blue-blue pairs. For red-blue pairs, the signal from the \citetalias{Lagos2012} SAM is similar to the observed one at small scales, but too high for $r>0.3\,h^{-1}\,\textrm{Mpc}$ .
    
    The difference between the SAMs is also visible in Table~\ref{tab:chi2}, whose upper part shows the $\chi^2_\textrm{redu}$ values for the different colour-selected lens pairs. We consider here $p=12$ data points and define a tension between observation and simulation at the 95\% confidence level (CL) if $\chi^2_\mathrm{redu}>1.75$. For the \citetalias{Henriques2015} SAM, $\chi^2_\textrm{redu}$ is smaller than this threshold for red-red, red-blue, and blue-blue lens pairs, so there is no tension between the observation and this model. The $\chi^2_\textrm{redu}$ for the \citetalias{Lagos2012} SAM, though, are notably higher than the threshold. Consequently, the predictions by the \citetalias{Lagos2012} SAM do not agree with the observations for these.

\begin{figure*}
        \centering
        \includegraphics[trim=25 25 70 65, clip,width=\linewidth]{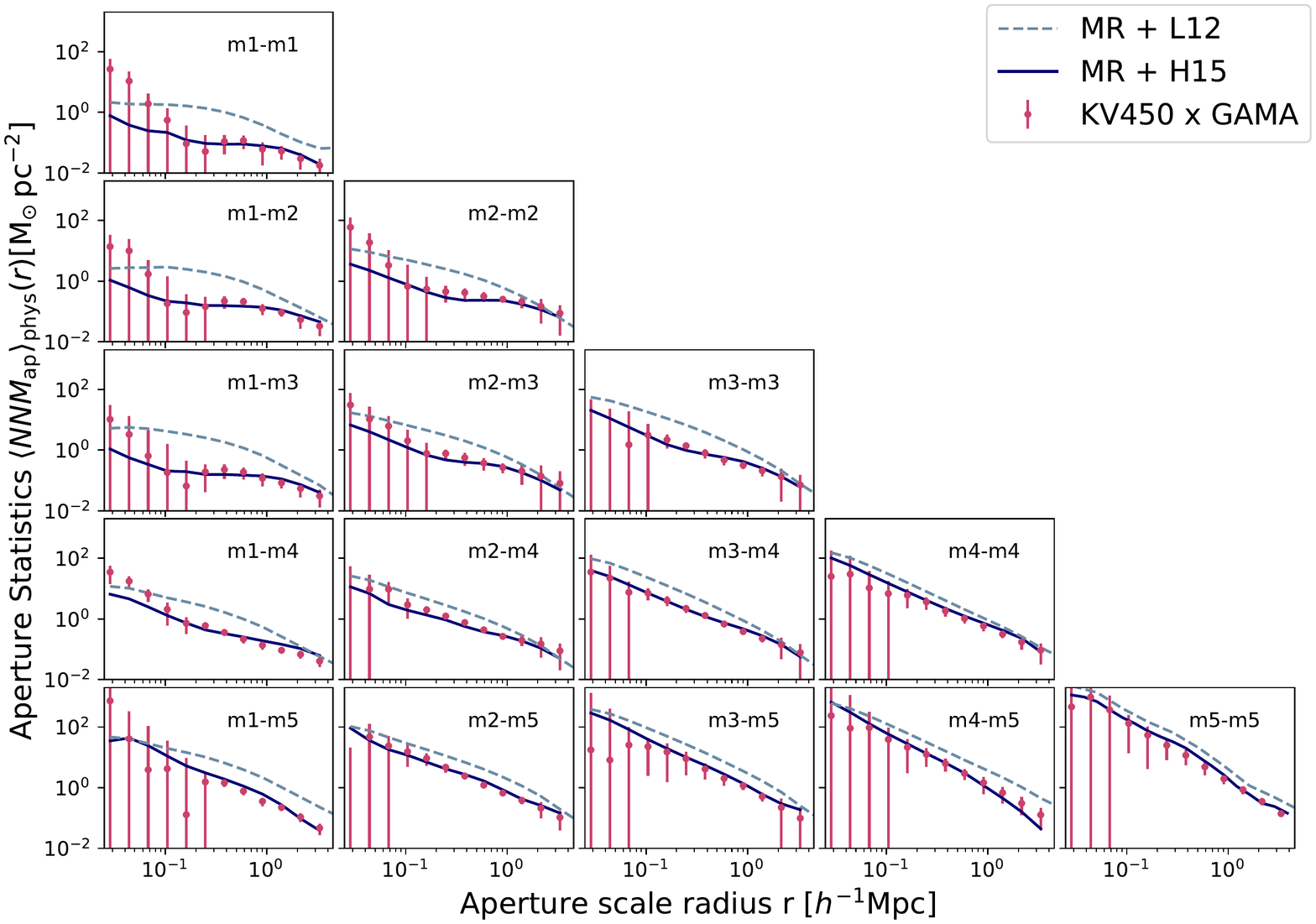}    
        \caption{Physical aperture statistics for stellar mass-selected lens samples in the MR with the \citetalias{Henriques2015} SAM (solid blue lines), the \citetalias{Lagos2012} SAM (dashed grey lines), and in GAMA with KV450 sources (pink points), using the mass bins defined in Table~\ref{tab:sub-samples}. Plots on the diagonal show the signal for unmixed lens pairs, while the other plots show the signal for mixed lens pairs. Error bars are the standard deviation from jackknife resampling.}
        \label{fig:NNMap_masses}
    \end{figure*}

    Figure~\ref{fig:NNMap_masses} shows the measured $\expval{\mathcal{NN} M_\mathrm{ap}}_\mathrm{phys}$ for lenses split by their stellar mass.
    The amplitude of the aperture statistics increases with the stellar mass of galaxies in a pair. Consequently, the bias factor increases with stellar mass. This trend exists for observed and both kinds of simulated lenses. Nevertheless, the predictions of the SAMs differ notably, with the aperture statistics obtained from the \citetalias{Lagos2012} SAM being substantially higher than those from the \citetalias{Henriques2015} SAM. The \citetalias{Lagos2012} SAM also deviates strongly from the observational measurements in KV450$\times$GAMA, that agree better with the \citetalias{Henriques2015} SAM. The deviation of the \citetalias{Lagos2012} SAM from the observations is strongest for lenses with $M^*\leq 10^{9.5}\,h^{-2}\mathrm{M}_{\odot}$ and decreases for larger stellar masses. 
    
    To quantify the deviation, we list the $\chi^2_\mathrm{redu}$ of the aperture statistics measured in the \citetalias{Henriques2015} and \citetalias{Lagos2012} SAM in the lower part of Table~\ref{tab:sn}. Again, a $\chi^2_\mathrm{redu}>1.75$ indicates a tension at the 95\% CL. The \citetalias{Lagos2012} SAM disagrees with the observation for all lens samples. The $\chi^2_\mathrm{redu}$ of the \citetalias{Henriques2015} SAM, though, are smaller than 1.75 for all but one correlation. The only tension exists for the correlation of lenses from stellar-mass samples m1 and m4, driven by differences at $r\lesssim0.2\,h^{-1}\,\mathrm{Mpc}$, where the \citetalias{Henriques2015} SAM underestimates $\expval{\mathcal{NN} M_\mathrm{ap}}_\textrm{phys}$.
    
    \begin{table}
        \caption{S/N of observed aperture statistics for the E-mode in the middle column and the B-mode in the right column}
        \label{tab:sn}
        \centering
        \begin{tabular}{ccccc}
            \hline\hline
            lens pairs & S/N of $\expval{\mathcal{NN} M_\mathrm{ap}}_\mathrm{phys}$ & S/N of $\expval{\mathcal{NN} M_{\perp}}$   \\
            \hline
            red -- red & 7.3 & 1.7 \\
            red -- blue & 5.1 & 1.4 \\
            blue -- blue & 4.7 & 0.6\\
            \hline
            m1 -- m1 & 3.9 & 1.7 \\
            m1 -- m2 & 4.1 & 1.4 \\
            m1 -- m3 & 3.7 & 0.7\\
            m1 -- m4 & 5.6 & 0.8\\
            m1 -- m5 & 9.2 & 0.7\\
            m2 -- m2 & 8.6 & 1.2\\
            m2 -- m3 & 3.3 & 1.8\\
            m2 -- m4 & 5.3 & 1.5\\
            m2 -- m5 & 3.4 & 0.9\\
            m3 -- m3 & 4.7 & 0.8\\
            m3 -- m4 & 5.7 & 0.9\\
            m3 -- m5 & 6.2 & 1.3\\
            m4 -- m4 & 7.1 & 0.7\\
            m4 -- m5 & 9.8 & 1.1\\
            m5 -- m5 & 9.1 & 0.4\\
            \hline
        \end{tabular}  
    \tablefoot{Samples are selected according to Table~\ref{tab:sub-samples}. B-mode is consistent with zero.}
    \end{table}
    
    Finally, we test for systematic effects by considering the B-mode $\expval{\mathcal{NN} M_{\perp}}$. Table~\ref{tab:sn} compares the S/Ns, defined by Eq.~\eqref{eq:S/N}, of $\expval{\mathcal{NN} M_\mathrm{ap}}_\textrm{phys}$ with those of the B-modes $\expval{\mathcal{NN} M_{\perp}}$ for all observed lens pairs. The S/Ns of $\expval{\mathcal{NN} M_{\perp}}$ are considerably smaller than the S/Ns of the $\expval{\mathcal{NN} M_\mathrm{ap}}_\textrm{phys}$, and they are consistent with a vanishing B-mode.

    
    \section{Discussion}
    \label{sec:discussion}
    We evaluated the SAM by \citetalias{Henriques2015} and the SAM by \citetalias{Lagos2012} by comparing their G3L predictions to measurements with KiDS, VIKING and GAMA. For this, we applied the improved estimator for the G3L three-point correlation function by \citetalias{Linke2020} and measured aperture statistics for mixed and unmixed lens galaxy pairs from colour- or stellar-mass-selected lens samples.
    
    Our measurements show a higher S/N than previous studies of G3L, due to the use of an improved estimator for G3L and new data. As shown in \citetalias{Linke2020}, redshift weighting increases the S/N of the aperture statistics by 35\% on mock data with a similar lens- and redshift distribution as in our observation. We also extended the considered scales. Therefore, we could probe the predictions of the SAMs well inside of dark matter halos at lengths below $1\,h^{-1}\mathrm{Mpc}$. These ranges are particularly interesting for testing galaxy formation models because the principal variations between different SAMs are the assumptions on phenomena, whose effects are most substantial at small scales, such as star formation, stellar and AGN feedback and environmental processes \citep{Guo2016}.
    
    The aperture statistics are larger for red-red lens pairs than for red-blue or blue-blue lens pairs, which indicates that red galaxies have higher bias factors than blue galaxies. We also found that the bias factor increases with stellar mass. These results support the general expectation that redder and more massive galaxies have higher bias factors, which has been found in multiple studies \citep[e.g.][]{Zehavi2002, Sheldon2004, Simon2018, Saghiha2017}.
    
    The predictions by the \citetalias{Henriques2015} SAM for aperture statistics of colour-selected lens samples agree with the observations at the 95\% CL. The signal predicted by the \citetalias{Lagos2012} SAM, though, deviates significantly from the observed G3L signal, being too high for red-red and red-blue, and too low for blue-blue pairs. 
    
    This deviation could be due to an overproduction of red galaxies in massive halos by the \citetalias{Lagos2012} SAM. As shown by \citet{Watts2005}, the G3L signal increases if more lens pairs reside in massive halos, so the relatively high $\expval{\mathcal{NN} M_\mathrm{ap}}_\textrm{phys}$  indicates that in the \citetalias{Lagos2012} SAM massive halos contain too many pairs of red galaxies. This interpretation is supported by studies by \citet{Baldry2006} of the \citet{Bower2006} SAM, on which the \citetalias{Lagos2012} SAM is based. They compare the fraction of red galaxies in the SAM with observations by the SDSS and found that the SAM predicts too many red galaxies, especially in regions of high surface mass density.
    
    \citet{Font2008} accredite the overproduction of red satellite galaxies to excessive tidal interactions and ram pressure stripping in the \citetalias{Lagos2012} SAM. This process decreases the amount of gas in satellite galaxies inside halos and thereby inhibits their star formation. Consequently, the stripped galaxies become redder, so the fraction of red galaxies increases, while the number of blue galaxies decreases. This effect could explain the low aperture statistics for blue-blue lens pairs in the \citet{Lagos2012} SAM, as fewer blue galaxies remain inside massive halos. 
    
    The aperture statistics for the stellar-mass-selected samples measured in the observation agree with the \citetalias{Henriques2015} SAM at the 95\% CL except for one sample. This finding is consistent with the conclusion by \citet{Saghiha2017}, although their study is limited to angular scales between \ang{;1;} and \ang{;10;}, did not consider mixed lens pairs and had a lower S/N due to the effect of chance lens pairs. 
    
    The \citetalias{Henriques2015} SAM agrees with the observations at the 95\% CL for all but the correlation of m1 and m4 lens galaxies. This difference is driven mainly by a low signal by the SAM at scales below $0.2\,h^{-1}\mathrm{Mpc}$. At these scales, the SAM also gives lower predictions for $\expval{\mathcal{NN} M_\mathrm{ap}}$ than the observations for m1-m2, m1-m3, and m2-m2 lens pairs. This trend could indicate that the SAM underpredicts G3L at small scales for low stellar masses. A possible reason is the limited resolution of the MR. The MRs softening length is $5\,h^{-1}\mathrm{kpc}$, so its spatial resolution is in the order of tens of kiloparsec \citep{Vogelsberger2020}. Therefore, the difference between the aperture statistics in the \citetalias{Henriques2015} SAM and the observation at small scales might be due to the limited resolution.
    
    The \citetalias{Lagos2012} SAM disagrees with the observations for all considered stellar-mass samples at the 95\% CL, and its predicted signal is significantly larger. The tension increases for lenses with lower stellar mass and is more prominent at smaller scales. 
    
    This tension might be due to inaccurate stellar masses of the simulated lens galaxies. If the SAM assigns too low stellar masses, galaxies from a higher stellar mass bin are incorrectly assigned to a lower mass bin, for example into m2 instead of m3. The SAM then overestimates the aperture statistics, because the bias factors of galaxies with larger stellar masses are higher. The choice of initial mass function could cause different stellar-mass assignments by the SAMs. While the \citetalias{Henriques2015} SAM used the same initial mass function as the observations \citep{Chabrier2003}, the \citetalias{Lagos2012} SAM assumes the initial mass function by \citet{Kennicutt1983}. Therefore, the stellar masses of the observation and the \citetalias{Lagos2012} might be inconsistent with each other.
    
    Another cause for the tension of the \citetalias{Lagos2012} SAM with the observation could be an overproduction of satellite galaxies inside massive halos. This interpretation agrees with \citet{Saghiha2017}, who find that the satellite fraction and mean halo masses for the \citetalias{Lagos2012} SAM is higher than for the \citetalias{Henriques2015} SAM. The tension between the \citetalias{Lagos2012} SAM and the observation increases for lower stellar masses and smaller scales, indicating that especially galaxies with low stellar mass are overproduced by the SAM and that their fraction rises closer to the centre of their dark matter halo. An excess of galaxies with small stellar masses would be consistent with excessive galaxy interactions inside halos. This finding, therefore, fits with the interpretation of the high G3L signal for red-red lens pairs in the SAM as caused by excessive ram pressure stripping.
    
    We presented the first measurements of G3L for mixed lens pairs and used the aperture statistics for red-blue lens pairs to test the linear deterministic bias model. This bias model predicts that the aperture statistics for mixed lens pairs is the geometric mean of the signals for equal lens pairs. Our observational measurements are consistent with this prediction, although the signal is too noisy at scales below $0.2\,h^{-1}\,\textrm{Mpc}$ for meaningful constraints on the bias model.
    
    The aperture statistics for mixed lens pairs are also useful to constrain the correlations of different galaxy populations inside the same dark matter halos. For example, the measured aperture statistics for red-blue lens pairs indicate that lens galaxies of different samples co-populate the same halos, as the signal would decrease at sub-Mpc scales due to a vanishing 1-halo term. Modelling of mixed-pair G3L in the context of the halo model will provide further insights into the correlation of galaxy populations inside halos. In contrast, GGL, which is only sensitive to the mean number of lenses inside halos and hence blind to the way mixed lens pairs populate halos, cannot yield the same information.    
    
    A compelling future study would be investigating whether full hydrodynamical simulations predict G3L with the same accuracy as the \citetalias{Henriques2015} SAM. Such a study would complement previous comparisons of GGL in hydrodynamical simulations to observations, for example by \citet{Velliscig2017} for the EAGLE simulation to KiDS and GAMA data, or \citet{Gouin2019} for the Horizon-AGN simulation to CFHTLenS and the Baryon Oscillation Spectroscopic Survey. While these studies conclude that the GGL predictions of these simulations agree with the observations, the same is not necessarily true for G3L, which depends on the correlation of matter and galaxy pairs.

    \begin{acknowledgements}
        LL is a member of and received financial support for this research from the International Max Planck Research School (IMPRS) for Astronomy and Astrophysics at the Universities of Bonn and Cologne. CH acknowledges support from the European Research Council under grant number 647112, and support from the Max Planck Society and the Alexander von Humboldt Foundation in the framework of the Max Planck-Humboldt Research Award endowed by the Federal Ministry of Education and Research. H. Hildebrandt is supported by a Heisenberg grant of the Deutsche Forschungsgemeinschaft (Hi 1495/5-1) as well as an ERC Consolidator Grant (No. 770935). AK acknowledges support from Vici grant 639.043.512, financed by the Netherlands Organisation for Scientific Research (NWO). CS acknowledges support from the Agencia Nacional de Investigaci\'on y Desarrollo (ANID) through grant FONDECYT Iniciaci\'on 11191125. AHW is supported by a European Research Council Consolidator Grant (No. 770935).
        Based on data products from observations made with ESO Telescopes at the La Silla Paranal Observatory under programme IDs 177.A-3016, 177.A-3017, 177.A-3018, 179.A-2004, 298.A-5015. We also use products from the GAMA survey. GAMA is a joint European-Australasian project based around a spectroscopic campaign using the Anglo-Australian Telescope. The GAMA input catalogue is based on data taken from the Sloan Digital Sky Survey and the UKIRT Infrared Deep Sky Survey. Complementary imaging of the GAMA regions is being obtained by several independent survey programmes including GALEX MIS, VST KiDS, VISTA VIKING, WISE, Herschel-ATLAS, GMRT and ASKAP providing UV to radio coverage. GAMA is funded by the STFC (UK), the ARC (Australia), the AAO, and the participating institutions. The GAMA website is http://www.gama-survey.org/. \\        
        \emph{Author contributions.} All authors contributed to the development and writing of this paper. The authorship list is given in two groups: The lead authors (LL, PSi, PS), followed by an alphabetical list of contributors to either the scientific analysis or the data products.
    \end{acknowledgements}

    \bibpunct{(}{)}{;}{a}{}{,}
    \bibliographystyle{aa} 
    \bibliography{biblio} 
    
    
    \begin{appendix}
        \section{Results for aperture statistics in angular units}
        \label{app:NNMap}
        
\begin{figure}
            \includegraphics[width=\linewidth, trim=0.2cm 0 1.4cm 0, clip]{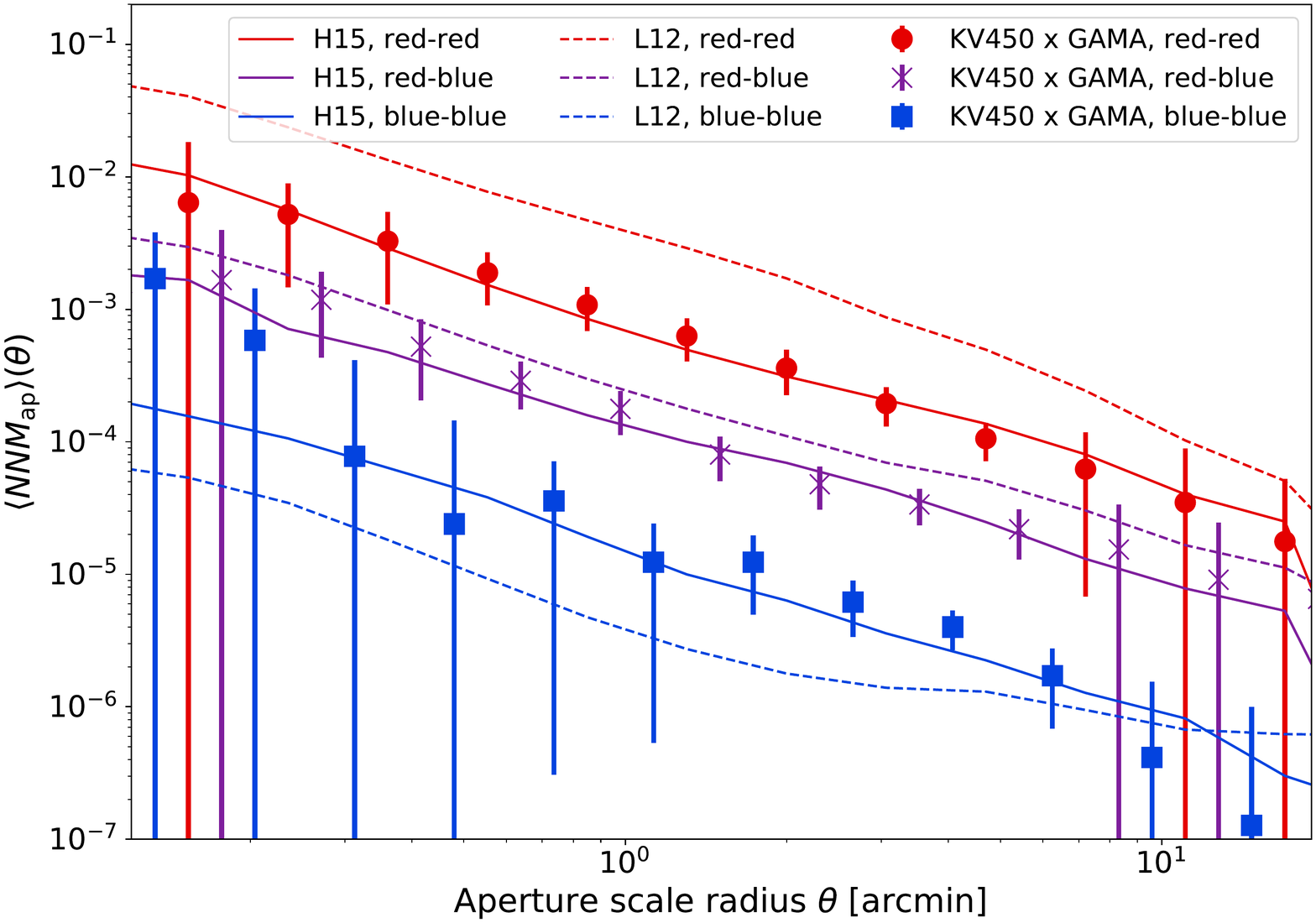}
            \caption{Aperture statistics in angular units for colour-selected lens samples of the \citetalias{Henriques2015} galaxies (solid lines), \citetalias{Lagos2012} galaxies (dashed lines) and KV450 $\times$ GAMA (points). The signal is shown for red-red lens pairs (red lines and filled circles), red-blue lens pairs (purple lines and crosses), and blue-blue lens pairs (blue lines and squares). Error bars on the observational measurements are the standard deviation from jackknifing.}
            \label{fig:NNMap_colours_angular}
        \end{figure}

\begin{figure*}
            \includegraphics[trim=25 30 70 65, clip,width=\linewidth]{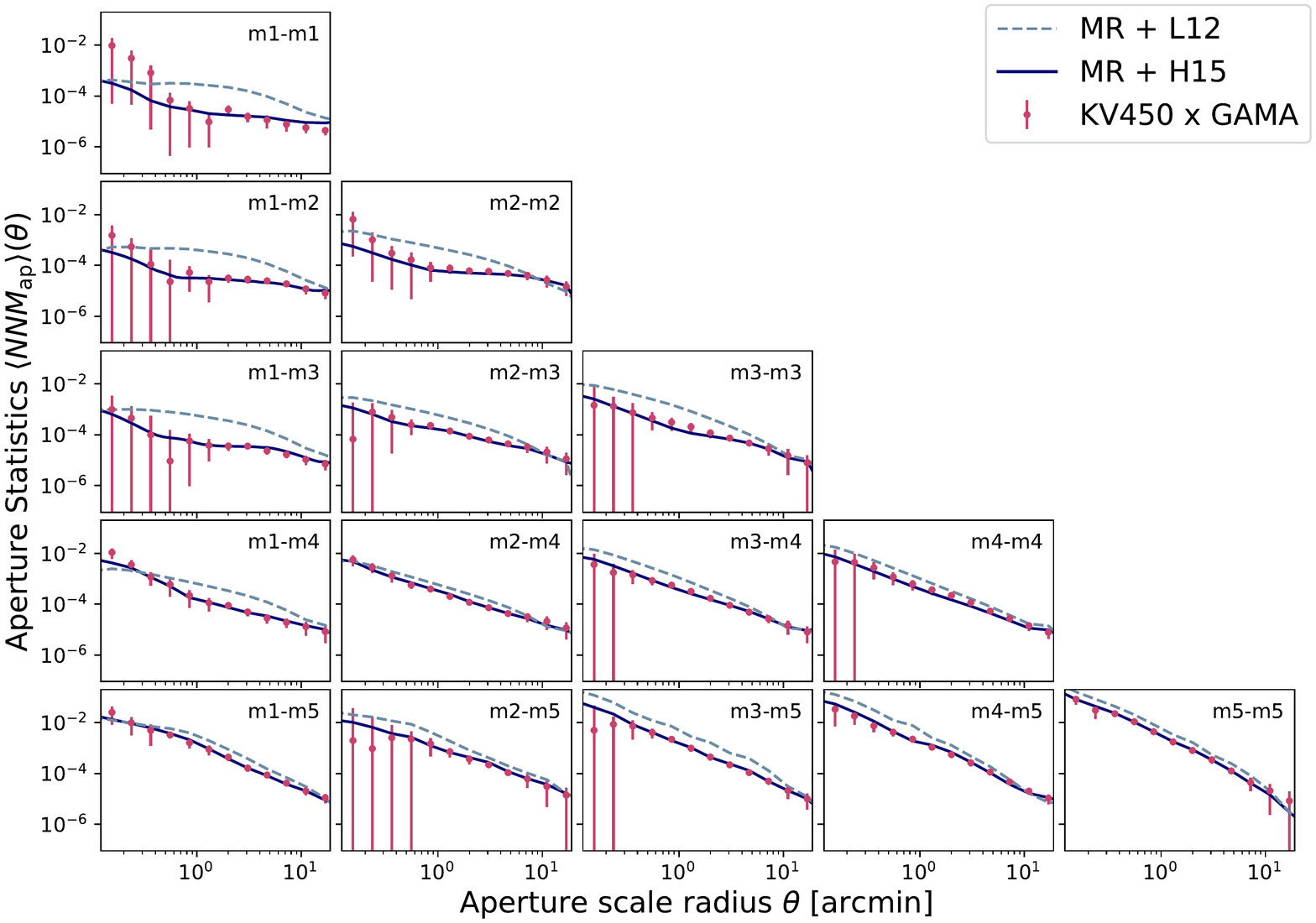}
            \caption{Angular aperture statistics for stellar mass-selected lens samples in the MR with the \citetalias{Henriques2015} SAM (solid blue lines), the \citetalias{Lagos2012} SAM (dashed grey lines), and in GAMA with KV450 sources (pink points), using the mass bins defined in Table~\ref{tab:sub-samples}. Plots on the diagonal show the signal for unmixed lens pairs, while the other plots show the signal for mixed lens pairs. Error bars are the standard deviation from jackknife resampling.}
            \label{fig:NNMap_masses_angular}
        \end{figure*}

        For completeness, we show here our results for the angular aperture statistics $\expval{\mathcal{NN} M_\mathrm{ap}}$, for colour-selected lens samples (Fig.~\ref{fig:NNMap_colours_angular}) and stellar-mass-selected lens samples (Fig.~\ref{fig:NNMap_masses_angular}). The $\expval{\mathcal{NN} M_\mathrm{ap}}$ exhibit similar trends to the $\expval{\mathcal{NN} M_\mathrm{ap}}_\textrm{phys}$ (see Sect~\ref{sec:results}). In particular, $\expval{\mathcal{NN} M_\mathrm{ap}}$ also increases with the lenses stellar masses and is larger for red-red than for red-blue or blue-blue lens galaxies. Furthermore, the predictions by the \citetalias{Henriques2015} SAM agrees well with the observed $\expval{\mathcal{NN} M_\mathrm{ap}}$, while the \citetalias{Lagos2012} SAM expects too large aperture statistics, especially for low stellar-mass galaxies. 
        
        The agreement of the \citetalias{Henriques2015} SAM and the discrepancy of the \citetalias{Lagos2012} SAM with the observations is supported by the $\chi^2_\textrm{redu}$ of the SAMs predictions for $\expval{\mathcal{NN} M_\mathrm{ap}}$, presented in Table~\ref{tab:chi2_angular}. The \citetalias{Henriques2015} SAM disagrees with the observations only for the correlation of m1 and m4 galaxies at the 95\% CL, while the \citetalias{Lagos2012} SAM is in tension with the observation for all samples.
        
        Note, that while the measurements of $\expval{\mathcal{NN} M_\mathrm{ap}}$ do not depend on the choice of cosmology, they change with the lens redshift distribution. Comparing $\expval{\mathcal{NN} M_\mathrm{ap}}$ measured in different observational surveys requires, therefore, careful consideration of the survey's selection functions.
        
        \begin{table}
            \caption{$\chi^2_\mathrm{redu}$ of $\expval{\mathcal{NN} M_\mathrm{ap}}$ for \citetalias{Henriques2015} and \citetalias{Lagos2012} SAMs.}
            \label{tab:chi2_angular}
            \centering
            \begin{tabular}{ccccc}
                \hline\hline
                lens pairs  & $\chi^2_\textrm{redu}$ for \citetalias{Henriques2015} & $\chi^2_\textrm{redu}$ for \citetalias{Lagos2012}\\
                \hline
                red -- red & 1.33 & \textbf{32.4} \\
                red -- blue & 0.39 & \textbf{1.92}\\
                blue -- blue & 0.85 & \textbf{2.31}\\
                \hline
                m1 -- m1 & 0.95 & \textbf{27.0}\\
                m1 -- m2 &  0.81 & \textbf{28.9}\\
                m1 -- m3 &  1.27 & \textbf{50.3}\\
                m1 -- m4 &  \textbf{3.69} &  \textbf{22.13}\\
                m1 -- m5 &  1.18 &  \textbf{5.29}\\
                m2 -- m2 &  1.29 &  \textbf{10.28}\\
                m2 -- m3 &  0.74 &  \textbf{17.13}\\
                m2 -- m4 &  0.45 &  \textbf{7.90}\\
                m2 -- m5 &  1.37 &  \textbf{21.66}\\
                m3 -- m3 &  0.40 &  \textbf{60.61}\\
                m3 -- m4 &  0.56 &  \textbf{18.57}\\
                m3 -- m5 &  0.90 &  \textbf{27.14}\\
                m4 -- m4 &  0.66 &  \textbf{3.15}\\
                m4 -- m5 &  1.36 &  \textbf{11.43}\\
                \hline
            \end{tabular}
            \tablefoot{Bold values indicate a tension at the 95\% CL.}
        \end{table}

    \end{appendix}

\end{document}